\documentstyle[12pt,epsfig]{article}
\def\be{\begin{equation}} \def\ee{\end{equation}} \def\bea{\begin{eqnarray}}
\def\eea{\end{eqnarray}} \def\nnb{\nonumber}

\begin{document}

\hfill{March 26, 2023,\ \ \ {\tt a12Cwarsapw\_ver6}} 

\begin{center}
\vskip 6mm 
\noindent
{\Large\bf  
The $S$ matrices of 
elastic $\alpha$-$^{12}$C scattering at low energies
in effective field theory
}
\vskip 6mm 

\noindent
{\large 
Shung-Ichi Ando\footnote{mailto:sando@sunmoon.ac.kr}, 
\vskip 6mm
\noindent
{\it
Department of Display and Semiconductor Engineering, \\
Sunmoon University, Asan, Chungnam 31460,
Republic of Korea
}
}
\end{center}

\vskip 6mm

The elastic $\alpha$-$^{12}$C scattering 
at low energies  
for $l=0,1,2,3,4,5,6$ 
is studied in effective field theory. 
We discuss the construction of the $S$ matrices of elastic 
$\alpha$-$^{12}$C scattering in terms of the amplitudes of
sub-threshold bound and resonant states of $^{16}$O, which are calculated 
from the effective Lagrangian. 
The parameters appearing in the $S$ matrices are fitted to the phase shift 
data below the $p$-$^{15}$N breakup threshold energy, and we find 
that the phase shifts are well described within the theory. 

\vskip 5mm 
\noindent PACS(s): 
11.10.Ef, 
24.10.-i, 
25.55.-e, 
26.20.Fj  

\newpage
\vskip 2mm \noindent
{\bf 1. Introduction}

Radiative $\alpha$ capture on carbon-12, $^{12}$C($\alpha$,$\gamma$)$^{16}$O,
is a fundamental reaction in nuclear astrophysics, 
which determines the C/O ratio along with the triple-$\alpha$
process in the helium burning
process in stars~\cite{f-rmp84}.
The radiative capture rate at the Gamow peak energy, $E_G=0.3~\textrm{MeV}$,
in the stars is, however, difficult to measure in the experimental facilities
because of the Coulomb barrier. One needs to employ a theoretical model,
fit the parameters of the model to the experimental data measured at a few
MeV energy or larger, and extrapolate the reaction rate down to the 
Gamow peak energy, $E_G=0.3~\textrm{MeV}$. 
Over the last half-century, many experimental and theoretical studies 
have been carried out. See, e.g., 
Refs.~\cite{bb-npa06,chk-epja15,bk-ppnp16,detal-rmp17,hkvk-rmp20,a-epja21} 
for review. 

The experimental data of elastic $\alpha$-$^{12}$C scattering provide us 
important information about the energies and widths of resonant states 
of $^{16}$O at low energies, which are used 
to fix some parameters of theoretical models. 
The first integrated phase shift analysis of the elastic scattering
for $l=0,1,2,3,4,5,6$ was reported with the data taken at the 
Ruhr-Universit\"{a}t Bochum by Plaga et al in 1987~\cite{petal-npa87}.
The upgraded precise phase shift analysis for partial waves, $l=0,1,2,3,4,5,6$,
was reported with the data taken at the University of Notre Dame
by Tischhauser et al. in 2009~\cite{tetal-prc09} where the energy range 
of the $\alpha$ particle is
$2.6~\textrm{MeV} \le E_\alpha \le 6.62~\textrm{MeV}$; $E_\alpha$ is the 
$\alpha$ energy in the lab frame. 
In this work, we study the elastic $\alpha$-$^{12}$C scattering at low energies
by employing an effective field theory (EFT).

To construct an EFT, one first needs to introduce a large momentum scale,
$\Lambda_H$.
An EFT is constructed by using the relevant degrees 
of freedom at low energy, below the large momentum scale, $\Lambda_H$. 
Then, the theory
provides us a perturbative (derivative) expansion scheme in powers of 
$Q/\Lambda_H$ where $Q$ is a typical momentum scale in a reaction in question. 
The coefficients of effective Lagrangian are fixed by using experimental data,
though they, in principle, can be determined 
from its mother theory~\cite{w-79}. 
An EFT is constructed for few-body nuclear systems, known as pionless EFT,
in which the pions are regarded as irrelevant degrees of freedom,
and one has $\Lambda_H = m_\pi$ where $m_\pi$ is the pion mass~\cite{bvk-02}. 
In addition, the dibaryon fields which have the baryon number, $B=2$, are
introduced in the theory to expand the terms around 
the unitary limit~\cite{k-npb97,bs-npa01,ah-prc05}. 
This expansion scheme turns out to reproduce the effective 
range expansion~\cite{b-pr49}. 
Furthermore, one can naturally extend the formalism 
to the studies of reactions 
involving photo emission~\cite{cs-prc99,r-npa00,aetal-prc06}, 
$\beta$ emission~\cite{kr-npa99,bc-plb01,aetal-plb08}, 
and neutrino reactions~\cite{bck-prc01,ash-prc20}. 
In the previous studies, we constructed 
an EFT for the $^{12}$C($\alpha$,$\gamma$)$^{16}$O reaction at the 
Gamow-peak energy, $E_G=0.3$~MeV, and studied the elastic $\alpha$-$^{12}$C
scattering at low energies for $l=0,1,2,3$ with and without the 
sub-threshold bound states of $^{16}$O~\cite{a-prc18,a-epja16}.
We subsequently studied the $E1$ transition of 
the $^{12}$C($\alpha$,$\gamma$)$^{16}$O reaction~\cite{a-prc19} 
and the $\beta$-delayed $\alpha$-emission from $^{16}$N~\cite{a-epja21} 
in the EFT. 

The inclusion of a resonant state in EFT has been studied by many authors,
e.g., by Gelman~\cite{g-prc09} and 
by Habashi, Fleming, and van Kolck~\cite{hfvk-epja21};
one needs to sum the leading order interactions up to the infinite order
at the vicinity of resonant energy, where in most cases 
ones consider a resonant state and a background contribution.
In real situations, on the other hand, a number of resonant states
are involved.
In the previous work, we studied the elastic $\alpha$-$^{12}$C scattering
for $l=2$ including the sub-threshold $2_1^+$ state and two resonant $2_2^+$
and $2_3^+$ states of $^{16}$O, in which 
the $S$ matrix is constructed from the amplitudes of those sub-threshold 
and resonant states; the amplitudes are derived from the effective Lagrangian
and represented in terms of the effective range parameters~\cite{a-prc22}.   
In the present work, we apply the method to the study of 
the elastic $\alpha$-$^{12}$C scattering at low energies 
for $l=0,1,2,3,4,5,6$. The parameters in the $S$ matrices are fitted 
to the phase shift data below the $p$-$^{15}$N breakup threshold energy,
and we find that the phase shifts are well described within the theory.
Then, we discuss the implication of the result for the application of EFT
to the study of nuclear reactions in stellar evolution. 

The present work is organized as the following. 
In section 2, an expression of the $S$ matrices is introduced and 
the effective Lagrangian is presented, and, in section 3, 
the elastic scattering 
amplitudes of the sub-threshold and resonant states for the $l$-th
partial wave states are derived from the Lagrangian.
In section 4, we discuss the numerical results of the parameter fit
to the phase shift data for $l=0,1,2,3,4,5,6$, and finally in section 5,
the results and discussion of this work are presented.

\vskip 2mm \noindent
{\bf 2. $S$ matrices and effective Lagrangian}

The construction of the $S$ matrix of the elastic $\alpha$-$^{12}$C scattering 
for $l=2$ was discussed in Ref.~\cite{a-prc22}. 
In this section, we review the method to extend it to 
the cases for $l$-th partial wave states.
The $S$ matrices of the elastic $\alpha$-$^{12}$C scattering for $l$-th 
partial wave channels are given as
\bea
S_l = e^{2i\delta_l}\,,
\label{eq;S_l}
\eea
where $\delta_l$ is the phase shift of elastic scattering 
for $l$-th partial wave channel whose experimental data
for $l=0,1,2,3,4,5,6$ at $2.6~\textrm{MeV} \le E_\alpha \le 6.62$~MeV 
are reported 
by Tischhauser et al.~\cite{tetal-prc09}.~\footnote{
The $\alpha$ energy labeled by $E_\alpha$ is in the lab frame,
and the other energies are given in the center-of-mass frame.
}
The scattering amplitude $\tilde{A}_l$
is related to the $S$ matrix as~\footnote{
There is a common factor difference between 
the expression of the amplitude $\tilde{A}_l$ and
the standard form of the amplitude $A_l$;  
$A_l = \frac{2\pi}{\mu}(2l+1) e^{2i\sigma_l} \tilde{A}_l$
where $\sigma_l$ is the Coulomb phase shift for $l$,
$e^{2i\sigma_l} = \Gamma(l+1+i\eta)/\Gamma(l+1-i\eta)$ with
$\eta=\kappa/p$.
} 
\bea
S_l = 1 + 2ip\tilde{A}_l\,.
\eea 

Various resonant states of $^{16}$O appear in the phase shift data,
and the sub-threshold bound states and resonant states at 
high energy (above the maximum energy of the data) may give contributions 
to the $S$ matrices of elastic $\alpha$-$^{12}$C scattering.
\begin{table}
\begin{center}
\begin{tabular}{c | c c c}
$l$ & (Bound states) & $2.6~\textrm{MeV} \le E_\alpha \le 6.62$~MeV, & 
6.62~MeV $< E_\alpha$ \cr \hline 
0 & $0_2^+$ & $0_3^+$ & $0_4^+$ \cr
1 & $1_1^-$ & $1_2^-$ & $1_3^-$ \cr
2 & $2_1^+$ & $2_2^+$, $2_3^+$  & $2_4^+$ \cr
3 & $3_1^-$ & $3_2^-$ & $3_3^-$ \cr
4 & --      & $4_1^+$, $4_2^+$ & $4_3^+$ \cr
5 & --      & --      & $5_1^-$ \cr
6 & (bg)    &  --     & $6_1^+$ \cr \hline
\end{tabular} 
\caption{
Bound and resonant ($l^\pi_{i-th}$) states of $^{16}$O,
which are used to construct the $S$ matrices
for $l$-th partial 
wave states of elastic $\alpha$-$^{12}$C scattering; 
the resonant states in the second column appear in the energy range,
$2.6~\textrm{MeV} \le E_\alpha \le 6.62$~MeV, and those in the third column 
do at $6.62~\textrm{MeV} < E_\alpha$. 
}
\label{table;states}
\end{center}
\end{table}
In Table \ref{table;states}, we present a list of the sub-threshold 
bound states, the resonant states appearing in the phase shift data, 
and the resonant states as background contributions 
from high energy
for the partial wave states of $\alpha$-$^{12}$C system.
By employing those states in the table,
we construct the $S$ matrices of the elastic $\alpha$-$^{12}$C scattering.

To construct an $S$ matrix, $S_l$, 
we may decompose a phase shift $\delta_l$,
for example, in the case of a sub-threshold bound state
and two resonant states, as~\cite{g-prc09}
\bea
\delta_l = \delta_l^{(bs)} + \delta_l^{(rs1)} 
+ \delta_l^{(rs2)}
\,,
\eea
where $\delta_l^{(bs)}$ is a phase shift from a sub-threshold 
bound state, and $\delta_2^{(rsN)}$ with $N=1,2$ 
are those from resonant states.
We now assume that each of those phase shifts may have 
a relation to a corresponding scattering amplitude
as 
\bea
e^{2i\delta_l^{(ch)}} &=& 1 + 2ip\tilde{A}_l^{(ch)}\,,
\eea
where $ch(annel) = bs, rs1, rs2$, 
and $\tilde{A}_l^{(bs)}$ and $\tilde{A}_l^{(rsN)}$ with $N=1,2$
are the amplitudes for the binding part and the first and second resonant 
parts of the amplitudes, 
which will be constructed from the effective Lagrangian 
in the next section. 
Thus, the total amplitude $\tilde{A}_l$ for the nuclear reaction part  
in terms of the three amplitudes, 
$\tilde{A}_l^{(bs)}$ and $\tilde{A}_l^{(rsN)}$ 
with $N=1,2$, is 
\bea
\tilde{A}_l &=&
\tilde{A}_l^{(bs)} 
+ e^{2i\delta_l^{(bs)}} \tilde{A}_l^{(rs1)} 
+ e^{2i(\delta_l^{(bs)}+\delta_l^{(rs1)})} \tilde{A}_l^{(rs2)} 
\,.
\label{eq;tldAl}
\eea
We note that the total amplitudes, $\tilde{A}_l$, are not obtained as
the summation of the amplitudes, $\tilde{A}_l^{(bs)}$ and 
$\tilde{A}_l^{(rsN)}$ with $N=1,2$, but have the additional phase factors
to $\tilde{A}_l^{(rsN)}$ with $N=1,2$, 
and the order of the three amplitudes, 
$\tilde{A}_l^{(bs)}$ and $\tilde{A}_l^{(rsN)}$ with $N=1,2$ are exchangeable.   

To study the elastic $\alpha$-$^{12}$C scattering at low energies in EFT,
we choose the $p$-$^{15}$N breakup threshold energy as the high energy scale.
For the relevant degrees of freedom of the theory, 
the ground $0^+$ states of $\alpha$ and $^{12}$C are chosen 
as elementary-like scalar fields.
We also introduce the composite fields of $\alpha$ and $^{12}$C to describe
the sub-threshold and resonant states of $^{16}$O.
An effective Lagrangian to derive the scattering amplitude
for the $l$-th wave elastic $\alpha$-$^{12}$C scattering 
at low energies including $l^\pi_{i-th}$ states of $^{16}$O 
may be written 
as~\cite{a-prc18,a-epja16}
\bea
{\cal L} &=& 
\phi_\alpha^\dagger \left(
iD_0 
+\frac{\vec{D}^2}{2m_\alpha}
\right) \phi_\alpha
+ \phi_C^\dagger\left(
iD_0
+ \frac{\vec{D}^2}{2m_C}
\right)\phi_C
\nnb \\ && +
\sum_{l=0}^6
\sum_{i}
\sum_{k=0}^3 
C_{(li)k}d_{(li)}^\dagger 
\left[
iD_0 
+ \frac{\vec{D}^2}{2(m_\alpha+m_C)}
\right]^k d_{(li)}
\nnb \\ && 
- 
\sum_{l=0}^6 
\sum_i
y_{(li)}\left[
d_{(nr)}^\dagger(\phi_\alpha O_{(l)}  \phi_C)
+ (\phi_\alpha O_{(l)}  \phi_C)^\dagger d_{(li)}
\right] \,,
\eea
where $\phi_\alpha$ ($m_\alpha$) and 
$\phi_C$ ($m_C$) are scalar fields (masses) of $\alpha$ and $^{12}$C, 
respectively.  
$D^\mu$ is a covariant derivative,
$D^\mu = \partial^\mu + i {\cal Q}A^\mu$ where
${\cal Q}$ is a charge operator and $A^\mu$ is the photon field.
$d_{(li)}$ are the composite fields for the $l^\pi_{i-th}$ states of  $^{16}$O 
consisting of $\alpha$ and $^{12}$C fields in $l$-th partial wave states, 
which are introduced 
for perturbative expansion around the unitary 
limit~\cite{k-npb97,bs-npa01,ah-prc05,b-pr49}.  
The field $d_{(li)}$ are tensors in general, which are represented as 
Cartesian tensors of rank $l$~\cite{s-93,j-prb70,ag-pra82} 
(we suppressed the indices of the Cartesian tensors); 
$O_{(l)}$ are also tensors to project the $\alpha$-$^{12}$C system 
to $l$-th partial wave states. 
The coupling constants, $C_{(li)k}$ with $k=0,1,2,3$, correspond to 
the effective range parameters of elastic $\alpha$-$^{12}$C
scattering; for those of the sub-threshold states of $^{16}$O, 
the first coupling constants, $C_{(li)k}$ with $k=0$, are fixed
by using the binding energies of the sub-threshold states 
and the other parameters are fitted to the experimental phase shift data
with other parameters appearing in the $S$ matrices.
For the coupling constants of the resonant parts, 
$C_{(li)k}$ with $k=0,1,2,3$, 
the first two terms, $C_{(li)k}$ with $k=0,1$,
are rewritten by using the resonant energies and widths of the 
resonant states of $^{16}$O, respectively. 
The third and fourth parameters for the resonant states, 
$C_{(li)k}$ with $k=2,3$ are fitted to the phase shift data.
The coupling constants $y_{(li)}$ are
convention-dependent~\cite{g-npa04}, and we take the convenient choice; 
$y_{(li)} =  \sqrt{2\pi(2l+1)\mu^{2l-1}}$
where $\mu$ is the reduced mass of $\alpha$ and 
$^{12}$C. 

\vskip 2mm \noindent
{\bf 3. Scattering amplitudes}

\begin{figure}[t]
\begin{center}
\resizebox{0.7\textwidth}{!}{
  \includegraphics{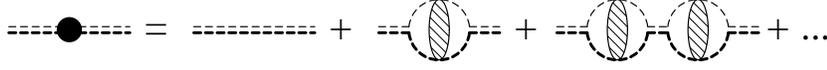}
}
\caption{
Diagrams for dressed $^{16}$O propagator.
A thick (thin) dashed line represents a propagator of $^{12}$C ($\alpha$),
and a thick and thin double dashed line with and without a filled circle
represent a dressed and bare $^{16}$O propagator, respectively.
A shaded blob represents
a set of diagrams consisting of all possible one-potential-photon-exchange
diagrams up to infinite order and no potential-photon-exchange one.
}
\label{fig;propagator}       
\end{center}
\end{figure}
\begin{figure}[t]
\begin{center}
\resizebox{0.2\textwidth}{!}{
 \includegraphics{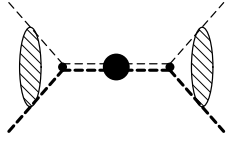}
}
\caption{
Diagram of the scattering amplitude.
See the caption of Fig.~\ref{fig;propagator} as well.
}
\label{fig;scattering_amplitude}      
\end{center}
\end{figure}

All the scattering amplitudes, $\tilde{A}_l^{(bs)}$ and 
$\tilde{A}_l^{(rsN)}$ with $N=1,2,3$, 
are calculated from the diagrams depicted in 
Figs.~\ref{fig;propagator} and \ref{fig;scattering_amplitude}.
The shaded blobs in the diagrams represent the parts of the non-perturbative
Coulomb interaction, the Coulomb propagator in the one-loop diagrams and
the Coulomb wavefunctions for the initial and final state interactions. 
Here, the bubble diagrams are summed up to infinite order 
in Fig.~\ref{fig;propagator}.
For the bound parts of the amplitudes, 
we treat them non-perturbatively
though they can be expanded perturbatively
at the energy region of the experimental data. 
For the resonant parts of the amplitudes
(for our case, they are classified as narrow resonances 
because of $\Gamma_r \ll E_r$~\cite{hfvk-epja21}), 
the counting rules of resonant states are carefully studied by 
Gelman~\cite{g-prc09} and 
Habashi, Fleming, and van Kolck~\cite{hfvk-epja21}.
The energy range of phase shift
data covers the resonant states, and at the vicinities of the resonant
energies we should have the amplitudes for which the bubble diagrams 
are summed up to the infinite order. 
While at the off-resonant energy regions, 
one can expand the resonant amplitudes perturbatively and the
$d_{(li)}$ fields may start mixing through the bubble diagram 
for corrections at higher orders. 
We keep the summed amplitudes for the resonant states 
as leading contributions
and ignore the field mixing. 

For the bound state amplitudes, $\tilde{A}_l^{(bs)}$ with $l=0,1,2,3$, 
we have~\cite{a-prc18,a-epja16} 
\bea
\tilde{A}^{(bs)}_l &=&
\frac{C_\eta^2 W_l(p)}{
K_l(p) 
-2\kappa H_l(p)
}
\label{eq;Aer2}
\label{eq;A2_nr}
\,,
\eea
where the function $C_\eta^2 W_l(p)$ in the numerator of the amplitude
is calculated from the initial and final state Coulomb interactions 
in Fig.~\ref{fig;scattering_amplitude}; $p$ is the magnitude of 
relative momentum of the $\alpha$-$^{12}$C system in the center of mass
frame, $p=\sqrt{2\mu E}$, and one has
\bea
C_\eta^2 &=& \frac{2\pi\eta}{\exp(2\pi\eta)-1}\,,
\\
W_l(p) &=& \left(\frac{\kappa^2}{l^2} + p^2\right) W_{l-1}(p)\,,
\ \ \ W_0(p) = 1\,,
\eea
where $\eta=\kappa/p$: 
$\kappa$ is the inverse of the Bohr radius,
$\kappa = Z_\alpha Z_{12C} \alpha_E\mu$, 
where $Z_A$ is the number of protons
of the nuclei, $Z_\alpha=2$ and $Z_{12C}=6$, 
and $\alpha_E$ is the fine structure constant.
The function $-2\kappa H_l(p)$ in the denominator of the amplitude
is the Coulomb self-energy term which is calculated 
from the loop diagram in Fig.~\ref{fig;propagator}, and one has 
\bea
H_l(p) &=& W_l(p) H(\eta)\, ,
\ \ \ 
H(\eta) = \psi(i\eta) + \frac{1}{2i\eta} -\log(i\eta)\,,
\eea
where $\psi(z)$ is the digamma function. 
The nuclear interaction is represented 
in terms of the effective range parameters 
in the function $K_l(p)$ in
the denominator of the amplitude in Eq.~(\ref{eq;Aer2}).
As discussed in Ref.~\cite{a-prc18}, large and significant contributions
to the series of effective range expansion,
compared to that evaluated from the phase shift data at the lowest energy
of the data, $E_\alpha=2.6$~MeV, 
appear from the Coulomb self-energy term, $-2\kappa H_l(p)$. 
In order to subtract those contributions, 
we include the effective range terms up to
$p^6$ order for $l=0,1,2$ and those up to $p^8$ order for $l=3$ 
as counter terms. 
Thus, we have  
\bea
K_l(p) &=&
-\frac{1}{a_l}
+\frac12 r_lp^2
-\frac14 P_l p^4
+Q_l p^6
-R_l p^8\,,
\eea
where $a_l$, $r_l$, $P_l$, $Q_l$, $R_l$ are effective range parameters. 
(We note that $R_l=0$ for $l=0,1,2$.)

Now we fix a parameter among the five effective range parameters,
$a_l$, $r_l$, $P_l$, $Q_l$, and $R_l$ by using the condition that 
the inverse of the scattering amplitude $\tilde{A}_l^{(bs)}$ vanishes at 
the binding energy of the sub-threshold states of $^{16}$O. 
Thus, the denominator of the scattering amplitude,
\bea 
D_l(p) = K_l(p) -2\kappa H_l(p) \,,  
\label{eq;binding_pole}
\eea
vanishes at $p=i\gamma_l$ where $\gamma_l$ are the binding momenta
of the $0_2^+$, $1_1^-$, $2_1^+$, $3_1^-$ ($l^\pi_{i-th}$) states of $^{16}$O; 
$\gamma_l = \sqrt{2\mu B_{l}}$ where $B_l$ 
are the binding energies of the bound states of $^{16}$O
from the $\alpha$-$^{12}$C breakup threshold.
At the binding energies, we have the wavefunction normalization factors 
$\sqrt{Z_l}$ for the bound states of $^{16}$O in the dressed $^{16}$O 
propagators as
\bea
\frac{1}{D_l(p)} = \frac{Z_l}{E+B_l} + \cdots\,,
\eea
where the dots denote the finite terms at $E=-B_l$.
Thus, one has
\bea
\sqrt{Z_l} = \left(
\left|
\frac{dD_l(p)}{dE}
\right|_{E=-B_l}
\right)^{-1/2}=
\left(
2\mu
\left|
\frac{dD_l(p)}{dp^2}
\right|_{p^2=-\gamma_l^2}
\right)^{-1/2}\,.
\eea
The wavefunction normalization factor $\sqrt{Z_l}$ is multiplied to a 
reaction amplitude when a bound state appears in the initial or final state of the reaction. 

Using the condition, $D_l(i\gamma_l)=0$, 
we fix the effective range parameter
$a_l$ as
\bea
-\frac{1}{a_l} &=& 
\frac12\gamma_l^2 r_l
+\frac14\gamma_l^4 P_l
+\gamma_l^6Q_l
+\gamma_l^8R_l 
+ 2\kappa H_l(i\gamma_l)\,. 
\label{eq;a2}
\eea
Using the relation of Eq.~(\ref{eq;a2}),
we rewrite the denominator of the amplitude $D_l(p)$ as
\bea
D_l(p) &=& 
\frac12 r_l (\gamma_l^2+p^2)
+\frac14 P_l (\gamma_l^4-p^4)
+ Q_l (\gamma_l^6 + p^6)
+ R_l (\gamma_l^8 - p^8)
\nnb \\ &&
+ 2\kappa \left[ H_2(i\gamma_l) - H_l(p)
\right]\,,
\eea
where we have three constants, $r_l$, $P_l$, $Q_l$
for $l=0,1,2$ and four constants, $r_3$, $P_3$, $Q_3$, $R_3$ 
for $l=3$ in the function $D_l(p)$ 
for the non-resonant amplitude $\tilde{A}_l^{(bs)}$,
which are fitted to the phase shift data. 
For the study of the asymptotic normalization coefficients (ANC),
the exponential factors in Eq.~(\ref{eq;tldAl})
almost become ones 
at the small energy region 
due to the Gamow factor in $C_\eta^2$, and the amplitudes
become
\bea
\tilde{A}_l &=& \tilde{A}_l^{(bs)} + \tilde{A}_l^{(rs1)} + \tilde{A}_l^{(rs2)}
+ O(C_\eta^4)\,,
\eea 
where the poles at the sub-threshold bound states of $^{16}$O exist in 
$\tilde{A}_l^{(bs)}$, the ANCs $|C_b|$ for the sub-threshold bound states
of $^{16}$O are calculated by using the formula~\cite{irs-prc84}
\bea
|C_b| = \frac{\gamma_l^l}{l!} \Gamma(l+1+\kappa/\gamma_l) 
\left( \left| \frac{dD_l(p)}{dp^2}\right|_{p^2 = - \gamma_l^2} 
\right)^{-1/2} \ \ (\textrm{fm}^{-1/2})\,,
\label{eq;Cb}
\eea
where $\Gamma(x)$ is the gamma function, and one may notice that the ANCs
are proportional to the wavefunction normalization factor $\sqrt{Z_l}$.
We note that the ANCs themselves are not necessary for the calculations of EFT
(while they may become constraints when the experimental data are not available for fitting 
the effective range parameters). 
In the present work, we display the values of the ANCs (in the next section)
as a demonstration
of the comparison with those obtained in our previous works as well as the other
theoretical models.  
In addition, we perform a test calculation to study the values of ANCs 
from a potential model. Its results are presented in the appendix. 

For the elastic scattering amplitudes for the resonant 
states of $^{16}$O, 
we may first have those amplitudes as the same expression 
of the bound state amplitudes in Eq.~(\ref{eq;Aer2})
in terms of the effective range parameters as
\bea
\tilde{A}^{(rsN)}_l &=&
\frac{C_\eta^2W_l(p)}{
K^{(rsN)}_l(p)
-2\kappa H_l(p)
}
\label{eq;A2_rs}
\,,
\eea
with $N=1,2,3$, which correspond to the first, second, and third 
resonant states of $^{16}$O, respectively, for $l$-th partial wave states and 
\bea
K^{(rsN)}_l(p) &=&
-\frac{1}{a_l^{(rsN)}}
+\frac12 r_l^{(rsN)} p^2
-\frac14 P_l^{(rsN)}  p^4
+Q_l^{(rsN)} p^6\,,
\label{eq;K2rsN}
\eea
where we include the terms up to $p^6$ order for all the resonant amplitudes.   

We now introduce 
the Taylor expansion around the resonant energies in the denominator
of the scattering amplitudes~\cite{hhvk-npa08}.
Thus, we rewrite the amplitudes as  
\bea
\tilde{A}_l^{(rsN)} &=&
- 
\frac{1}{p}
\frac{\frac12\Gamma_{(li)}(E) }{E-E_{R(li)} 
+ R_{(li)}(E) + i\frac12\Gamma_{(li)}(E)}\,,
\label{eq;A2_rsN}
\eea
with
\bea
\Gamma_{(li)}(E) &=& \Gamma_{R(li)}
\frac{pC_\eta^2W_l(p)}
     {p_rC_{\eta_r}^2W_l(p_r)}\,,
\\
R_{(li)}(E) &=& a_{(li)}(E-E_{R(li)})^2 + b_{(li)}(E-E_{R(li)})^3 
\,,
\label{eq;R}
\eea
where 
\bea
a_{(li)} &=& \frac12 Z_{R(li)} 
\left(
2P_l^{(rsN)} \mu^2
-48Q_l^{(rsN)} \mu^3 E_{R(li)}
+ 2\kappa Re 
\left. 
\frac{\partial^2 H_l}{\partial E^2}
\right|_{E=E_{R(li)}}
\right) \,,
\\
b_{(li)} &=& \frac16 Z_{R(li)}
\left(
-48Q_l^{(rsN)} \mu^3 
+ 2\kappa \left. Re
\frac{\partial^3 H_l}{\partial E^3}
\right|_{E=E_{R(li)}}
\right)\,,
\\
Z_{R(li)}^{-1} &=&
Re \left.
\frac{\partial}{\partial E} D_l^{(rsN)}(E)\right|_{E=E_{R(li)}}\,,
\ \ \ \
Z_{R(li)} = 
\frac{\Gamma_{R(li)}}{
2p_rW_l(p_r)C_{\eta_r}^2
}\,.
\eea
In the above equations,
$E_{R(li)}$  and $\Gamma_{R(li)}$ are 
the energies and the widths of the
resonant $l^\pi_{i-th}$ states of $^{16}$O, and $p_r$ are the resonant momenta,
$p_r=\sqrt{2\mu E_{R(li)}}$,  
which also appear in $\eta_r$ as $\eta_r=\kappa/p_r$.
We note that the expression of Eq.~(\ref{eq;A2_rsN}) resembles 
that of the Breit-Wigner formula, but it is derived from the expression
of the effective range expansion in Eq.~(\ref{eq;K2rsN}); 
it has the additional terms,
the corrections of the higher order terms being proportional to 
$(E-E_{R(li)})^2$ and  
$(E-E_{R(li)})^3$ in the function $R_{(li)}(E)$ in Eq.~(\ref{eq;R}); 
though the coefficients $a_{(li)}$ and $b_{(li)}$ are functions
of the effective range parameters, $P_l^{(rsN)}$ and $Q_l^{(rsN)}$,
we treat $a_{(li)}$ and $b_{li)}$ as independent free parameters
for the sake of simplicity.  

Using the expression of the amplitudes 
in Eqs.~(\ref{eq;A2_nr}) and (\ref{eq;A2_rsN}), 
we obtain an expression of the $S$ matrices in Eq.~(\ref{eq;S_l}) as
\bea
e^{2i\delta_l} &=& 
\frac{K_l(p) - 2\kappa Re H_l(p) + ipC_\eta^2W_l(p)}
     {K_l(p) - 2\kappa Re H_l(p) - ipC_\eta^2W_l(p)}
\prod_i
\frac{E - E_{R(li)} + R_{(li)}(E) - i\frac12\Gamma_{(li)}(E)}
     {E - E_{R(li)} + R_{(li)}(E) + i\frac12\Gamma_{(li)}(E)}
\,,
\label{eq;exp2idel_l} 
\eea
where the part for the bound states in terms of
$K_l(p)-2\kappa Re H_l(p) \pm ipC_\eta^2W_l(p)$ 
appear for $l=0,1,2,3$, and we also include it for $l=6$ as 
a background contribution from low energy. 
The part of the resonant states in terms of 
$E-E_{R(li)} + R_{(li)}(E) \mp i \frac12 \Gamma_{(li)}(E)$ 
are for those appearing in the energy range of the phase shift data, 
and we also include it for each of the partial waves as a background 
contribution from high energy.

\vskip 2mm \noindent
{\bf 4. Numerical results}

We construct an $S$ matrix of the elastic scattering 
for each of the partial waves, $l=0,1,2,3,4,5,6$, and fit 
parameters in the $S$ matrices to the phase shift data
at $2.6~\textrm{MeV} < E_\alpha < 6.62~\textrm{MeV}$
reported by Tischhauser et al.~\cite{tetal-prc09}, 
by means of a Markov chain Monte Carlo (MCMC) program~\cite{emcee}.  
We will see that curves calculated by using the fitted parameters
reproduce the phase shift data very well.

\vskip 2mm \noindent
{\bf 4.1  Phase shift for $l=0$ channel}

We consider three states, $0_2^+$, $0_3^+$, $0_4^+$ states of $^{16}$O
for the $S$ matrix of elastic $\alpha$-$^{12}$C scattering for $l=0$,
where $0_2^+$ is the sub-threshold bound state,
$0_3^+$ is the resonant state appearing in the phase shift data 
at $E_\alpha= 6.52$~MeV,
and $0_4^+$ is the resonant state as a background contribution from high
energy appearing at $E_\alpha=9.16$~MeV.
Thus, we have an expression of the $S$ matrix for $l=0$ as 
\bea
e^{2i\delta_0} &=& 
\frac{K_0(p) - 2\kappa Re H_0(p) + ip C_\eta^2}
     {K_0(p) - 2\kappa Re H_0(p) - ip C_\eta^2}
\prod_{i=3}^4
\frac{E-E_{R(0i)} + R_{(0i)}(E) - i\frac12\Gamma_{(0i)}(E)}
     {E-E_{R(0i)} + R_{(0i)}(E) + i\frac12\Gamma_{(0i)}(E)}
\,,
\label{eq;exp2idel0}
\eea
with
\bea
K_0(p) &=& 
\frac12 r_0(\gamma_0^2 + p^2)
+ \frac14P_0(\gamma_0^4 - p^4)
+ Q_0(\gamma_0^6 + p^6)
+ 2\kappa H_0(i\gamma_0)\,,
\\
\Gamma_{(0i)}(E) &=& \Gamma_{R(0i)}
\frac{pC_\eta^2W_0(p)}
     {p_rC_{\eta_r}^2W_0(p_r)}
\,,
\\
R_{(0i)}(E) &=& 
a_{(0i)} (E - E_{R(0i)})^2
+ b_{(0i)} (E - E_{R(0i)})^3\,,
\ \ \ i=3,4
\eea
where $\gamma_0=\sqrt{2\mu B_0}$ and $p_r =\sqrt{2\mu E_{(0i)}}$.
Thus, we have 7 parameters to fit the data,
\bea
\theta_0 &=& \{
r_0, P_0, Q_0, 
E_{R(03)}, \Gamma_{R(03)}, 
a_{(04)}, b_{(04)}
\}\,,
\eea 
where two parameters $a_{(03)}$ and $b_{(03)}$ are set to be zero,
$a_{(03)}=b_{(03)}=0$, because they are not sensitive to the parameter fit.
We use the experimental values for $E_{R(04)}$ and $\Gamma_{R(04)}$,
$E_{R(04)}^{(exp)} = 6.870(15)$~MeV and 
$\Gamma_{R(04)}^{(exp)}=185(35)$~keV~\cite{twc-npa93},
because they are not covered by the phase shift data, to which one cannot fit
them. 

\begin{table}[h]
\begin{center}
\begin{tabular}{c | c c c c c c} 
$l^\pi_{i-th}$ & $p^0~\textrm{order}$ & $p^2$ & 
 $p^4$ & $p^6$ & $p^8$ 
\cr \hline \hline
$0_2^+$ & $a_0(\textrm{fm})$ & $r_0$($\textrm{fm}$) & $P_0$($\textrm{fm}^3$) & 
$Q_0$($\textrm{rm}^5$) & 
\cr 
 & --- & 0.26847(1) & $-0.0363(4)$ & 0.0011(1) & 
\cr \hline
$0_3^+$  & $E_{R(03)}$(MeV) & $\Gamma_{R(03)}$(keV) &
\cr 
 & 4.8884(1) & 1.34(3) & 
\cr \hline 
$0_4^+$ & $E_{R(04)}(\textrm{MeV})$ & $\Gamma_{R(04)}(\textrm{keV})$ & 
 $a_{(04)}$($\textrm{MeV}^{-1}$) & 
 $b_{(04)}$($\textrm{MeV}^{-2}$) 
\cr 
 & --- & --- & 0.75(1) & 0.18(1) &  
\cr \hline 
\hline
$1_1^-$, $1_2^-$ & $a_1(\textrm{fm}^{3})$ & $r_1$($\textrm{fm}^{-1}$) & $P_1$($\textrm{fm}$) & 
$Q_1$($\textrm{fm}^3$) 
\cr 
 & --- & 0.415314(7) & $-0.57428(7)$ & 0.02032(2) 
\cr \hline
$1_3^-$ & $E_{R(13)}(\textrm{MeV})$ & $\Gamma_{R(13)}(\textrm{keV})$ & 
 $a_{(13)}$($\textrm{MeV}^{-1}$) & 
 $b_{(13)}$($\textrm{MeV}^{-2}$) 
\cr
 & --- & --- & 0.43(25) & 3.8(7)
\cr \hline 
\hline
$2_1^+$ & $a_2(\textrm{fm}^5)$ & $r_2$($\textrm{fm}^{-3}$) & 
 $P_2$($\textrm{fm}^{-1}$) &
$Q_2$($\textrm{fm}$) 
\cr 
 & --- & 0.149(4) & $-1.19(5)$ & 0.081(16) 
\cr \hline
$2_2^+$ & $E_{R(22)}$($\textrm{MeV}$) & $\Gamma_{R(22)}$($\textrm{keV}$) &
 & 
\cr
 & 2.68308(5) & 0.75(2) & 
\cr \hline
$2_3^+$ & $E_{R(23)}$(MeV) & $\Gamma_{R(23)}$(keV) &
$a_{(23)}$($\textrm{MeV}^{-1}$) & $b_{(23)}$($\textrm{MeV}^{-2}$) 
\cr
 & 4.3545(2) & 74.61(3) & 0.46(12) & 0.49(9)
\cr \hline 
$2_4^+$ & $E_{R(24)}(\textrm{MeV)}$ & $\Gamma_{R(24)}(\textrm{keV})$ & & 
\cr 
 & --- & --- & 
\cr \hline 
\hline
$3_1^-$, $3_2^-$ & $a_3(\textrm{fm}^7)$ & $r_3$($\textrm{fm}^{-5}$) & 
 $P_3$($\textrm{fm}^{-3}$) &
 $Q_3$($\textrm{fm}^{-1}$) & $R_3$($\textrm{fm}$) 
\cr 
 & --- & 0.0335(2) & $-0.446(9)$ & 0.311(5) & $-0.152(3)$ 
\cr \hline 
$3_3^-$ & $E_{R(33)}(\textrm{MeV})$ & $\Gamma_{R(33)}(\textrm{keV})$ &
$a_{(33)}(\textrm{MeV}^{-1})$ & $b_{(33)}(\textrm{MeV}^{-2})$ 
\cr
 & --- & --- & 32(32) & $3.2(32)\times 10^2$
\cr \hline 
\hline
$4_1^+$ & $E_{R(41)}$(MeV) & $\Gamma_{R(41)}$(keV) & 
$a_{(41)}$($\textrm{MeV}^{-1}$) & $b_{(41)}$($\textrm{MeV}^{-2}$) 
\cr 
 & 3.19606(1) & 25.91(1) & 0.740(3) & 0.304(5) 
\cr \hline
$4_2^+$ & $E_{R(42)}$(MeV) & $\Gamma_{R(42)}$(keV) & 
\cr
 & 3.93655(2) & 0.425(4) &  
\cr \hline
$4_3^+$ & $E_{R(41)}(\textrm{MeV})$ & $\Gamma_{R(41)}(\textrm{keV})$ &
$a_{(43)}$($\textrm{MeV}^{-1}$) & $b_{(43)}$($\textrm{MeV}^{-2}$) 
\cr
 & --- & ---  & 0.889(6) & 0.216(3) 
\cr \hline 
\hline
$5_1^-$ & $E_{R(51)}(\textrm{MeV})$ & $\Gamma_{R(51)}(\textrm{keV})$ &
 $a_{(51)}$($\textrm{MeV}^{-1}$) & $b_{(51)}$($\textrm{MeV}^{-2}$) 
\cr 
 & --- & --- & 0.572(6) & 0.104(2) 
\cr \hline 
\hline
(bg) & & $r_6$($\textrm{fm}^{-11}$) & $P_6$($\textrm{fm}^{-9}$) 
\cr
 & & $-0.3(2)$ & 2(1)
\cr \hline
$6_1^+$ & $E_{R(61)}(\textrm{MeV})$ & $\Gamma_{R(61)}(\textrm{keV)}$ &
  $a_{(61)}$($\textrm{MeV}^{-1}$) & $b_{(61)}$($\textrm{MeV}^{-2}$)
\cr
 & --- & --- & 0.8(1) & 0.18(4) 
\cr\hline 
\end{tabular}
\caption{
Fitted values of the parameters in the amplitudes of the 
sub-threshold and resonant
states of $^{16}$O (and a background contribution for $l=6$), 
which are listed in table \ref{table;states}, 
in the $S$ matrices of elastic $\alpha$-$^{12}$C
scattering for $l=0,1,2,3,4,5,6$. 
The parameters whose values are not shown in the table are fixed 
by using the experimental data. 
Parameters not shown in the table are not included in the 
parameter fit.  
}
\label{table;parameters_for_l=0123456}
\end{center}
\end{table}
\begin{table}[h]
\begin{center}
\begin{tabular}{c | c c c c c c c} 
$l$ & 0 & 1 & 2 & 3 & 4 & 5 & 6 
\cr \hline
$\chi^2/N$ & 0.013 & 0.089 & 0.66 & 0.87 & 0.47 & 0.094 & 0.026
\end{tabular}
\caption{Values of $\chi^2/N$ for $l=0,1,2,3,4,5,6$
for the parameter fit where $N=252$ for $l=0,1,2,3,4$, 
$N=243$ for $l=5$, and $N=186$ for $l=6$.  
}
\label{table;chi2_over_N}
\end{center}
\end{table}
\begin{table}[h]
\begin{center}
\begin{tabular}{c | c c c c } 
$l_{i-th}^\pi$ & $0_2^+$ & $1_1^-$ & $2_1^+$ & $3_1^-$  
\cr \hline
$|C_b| (\textrm{fm}^{-1/2})$ & 370(25) & $1.727(3)\times 10^{14}$ & 
 $3.1(6)\times 10^4$ & 113(8)  
\end{tabular}
\caption{Values of ANC, $|C_b|$,
for the sub-threshold $0_2^+$, $1_1^-$, $2_1^+$, $3_1^-$ states of 
$^{16}$O. 
}
\label{table;ANC}
\end{center}
\end{table}

In Table~\ref{table;parameters_for_l=0123456},
we show fitted values and their errors of the 7 parameters in the amplitudes
of the $0_2^+$, $0_3^+$, $0_4^+$ states of $^{16}$O in the $S$ matrix of 
elastic $\alpha$-$^{12}$C scattering for $l=0$ in Eq.~(\ref{eq;exp2idel0}) 
where we find a small value of $\chi^2/N$, $\chi^2/N = 0.013$, for the 
parameter fit, as shown in Table~\ref{table;chi2_over_N}.
The fitted values of $E_{R(03)}$ and $\Gamma_{R(03)}$ agree with 
their experimental values, $E_{R(03)}^{(exp)} = 4.887(2)$~MeV and 
$\Gamma_{R(03)}^{(exp)} = 1.5(5)$~keV~\cite{twc-npa93}. 
Those of the parameters, such as $P_0$, $a_{(03)}$, and $b_{(03)}$, 
at higher order have large errors.  
One may notice that the errors of the effective range parameters, $r_0$,
$P_0$, $Q_0$ become larger in order 
as the orders of the $p^2$ expansion increase. 
This may indicate that the perturbative expansion in the 
effective range parameters works well. 

In Table~\ref{table;ANC}, we show a value of the ANC, $|C_b|$, of
the $0_2^+$ state of $^{16}$O; we obtain $|C_b| = 370(25)~\textrm{fm}^{-1/2}$,
which is smaller than our previous estimates, 
$|C_b|= 443(3)~\textrm{fm}^{-1/2}$~\cite{a-prc18}
and $|C_b|=(6.4\textrm{--}7.4)\times 10^2~\textrm{fm}^{-1/2}$~\cite{a-prc20}.
We note that the values of ANC of the $0_2^+$ state of $^{16}$O reported
in the literature are still scattered: from the $R$-matrix analysis, 
the reported values of ANC are 
$|C_b|=44^{+270}_{-40}~\textrm{fm}^{-1/2}$~\cite{setal-plb11},
1800$~\textrm{fm}^{-1/2}$~\cite{detal-prc13}, 
and 1560$~\textrm{fm}^{-1/2}$~\cite{detal-rmp17},
from the $\alpha$ transfer reaction, 
1560(100)$~\textrm{fm}^{-1/2}$~\cite{aetal-prl15},
and from those fitting the phase shift data using the square-well potential, 
3218.46$~\textrm{fm}^{-1/2}$~\cite{bkms-prc18} 
and 886 -- 1139$~\textrm{fm}^{-1/2}$~\cite{bkms-22}, and 
using the so-called $\Delta$-method, 406$~\textrm{fm}^{-1/2}$~\cite{oin-prc17} 
and 293$~\textrm{fm}^{-1/2}$~\cite{o-npa21}. 
So, this would be an interesting issue to investigate in the future.   

\begin{figure}[h]
\begin{center}
  \includegraphics[width=12cm]{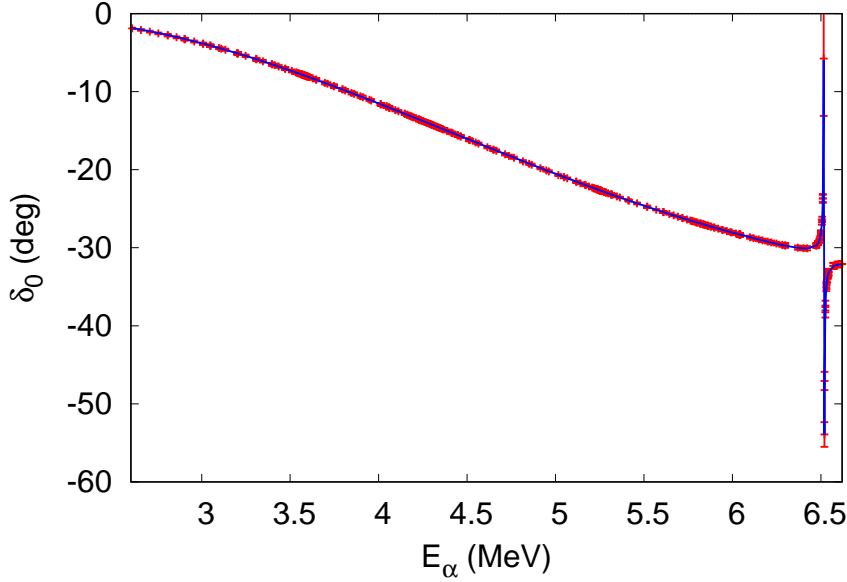}
\caption{
Phase shift $\delta_0$ of elastic $\alpha$-$^{12}$C scattering
for $s$-wave channel 
as a function of 
$E_\alpha$ calculated by using the fitted values of the parameters. 
Experimental data are included in the figure as well. 
}
\label{fig;del0}       
\end{center}
\end{figure}
In Fig.~\ref{fig;del0}, we plot a curve of the phase shift for $l=0$, 
$\delta_0$, by using the fitted values of the parameters obtained in 
Table~\ref{table;parameters_for_l=0123456}. The phase shift data are also
displayed in the figure.
We find that the fitted curve agrees well with the phase shift data.

\vskip 2mm \noindent
{\bf 4.2  Phase shift for $l=1$ channel}

We consider three states, $1_1^-$, $1_2^-$, $1_3^-$ states of $^{16}$O 
to construct the $S$ matrix of elastic $\alpha$-$^{12}$C scattering for
$l=1$ where $1_1^-$ is the sub-threshold bound state, $1_2^-$ is the 
resonant state appearing in the phase shift data at $E_\alpha= 3.23$~MeV, 
and $1_3^-$ is a resonant state as a background contribution from high energy
appearing at $E_\alpha = 7.04$~MeV.  
Because the resonant $1_2^-$ state can be described by the 
effective range parameters for the sub-threshold bound $1_1^-$ state, 
as discussed in Ref.~\cite{a-prc19},
we include the $1_2^-$ state in the amplitude of the 
sub-threshold $1_1^-$ state.
Thus, we have an expression of the $S$ matrix for $l=1$ as
\bea
e^{2i\delta_1} &=& 
\frac{K_1(p) -2\kappa Re H_1(p) + ipC_\eta^2 W_1(p)}
     {K_1(p) -2\kappa Re H_1(p) - ipC_\eta^2 W_1(p)}
\frac{E - E_{R(13)} + R_{(13)}(E) - i\frac12\Gamma_{(13)}(E)}
     {E - E_{R(13)} + R_{(13)}(E) + i\frac12\Gamma_{(13)}(E)}
\,,
\label{eq;exp2idel1}
\eea
with
\bea
K_1(p) &=& 
\frac12r_1(\gamma_1^2 + p^2)
+ \frac14P_1(\gamma_1^4 - p^4)
+ Q_1 (\gamma_1^6 + p^6)
+ 2\kappa H_1(i\gamma_1)\,,
\\
\Gamma_{(13)}(E) &=& \Gamma_{R(13)} 
\frac{pC_\eta^2 W_1(p)}
     {p_rC_{\eta_r}^2 W_1(p_r)}
\,,
\\
R_{(13)}(E) &=& 
a_{(13)}(E - E_{R(13)})^2 
+ b_{(13)}(E - E_{R(13)})^3\,, 
\eea
where $\gamma_1 = \sqrt{2\mu B_1}$ and $p_r = \sqrt{2\mu E_{R(13)}}$.
We have 5 parameters to fit the data as
\bea
\theta_1 = \{
r_1, P_1, Q_1, a_{(13)}, b_{(13)}
\}\,,
\eea
and we use the experimental values for $E_{R(13)}$ and $\Gamma_{R(13)}$,
$E_{R(13)}^{(exp)} = 5.278(2)$~MeV and 
$\Gamma_{R(13)}^{(exp)} =91(6)$~keV~\cite{twc-npa93},
for the background contribution from high energy. 

In Table~\ref{table;parameters_for_l=0123456},
we show fitted values and their errors of the 5 parameters in the amplitudes
of the $1_1^-$, $1_2^-$, $1_3^-$ states of $^{16}$O in the $S$ matrix of 
elastic $\alpha$-$^{12}$C scattering for $l=1$ in Eq.~(\ref{eq;exp2idel1})
where we find a small value of $\chi^2/N$, $\chi^2/N = 0.089$, for the 
parameter fit, as shown in Table~\ref{table;chi2_over_N}. 
We also find relatively large error bars of the coefficients $a_{(13)}$ and
$b_{(13)}$ at the high order while all errors of the effective range
coefficients, $r_1$, $P_1$, $Q_1$, turn out to be small presumably 
because those parameters need to fit the resonant $1_2^-$ state of $^{16}$O 
appearing in the phase shift data. 
In Table~\ref{table;ANC}, we show the value of the ANC of the sub-threshold 
$1_1^-$ state of $^{16}$O; we have 
$|C_b|=1.727(3)\times 10^{14}~\textrm{fm}^{-1/2}$ which agrees well to 
our previous estimate, 
$|C_b| = (1.6\textrm{--}1.9)\times 10^{14}~\textrm{fm}^{-1/2}$~\cite{a-prc18}.
\begin{figure}[h]
\begin{center}
  \includegraphics[width=12cm]{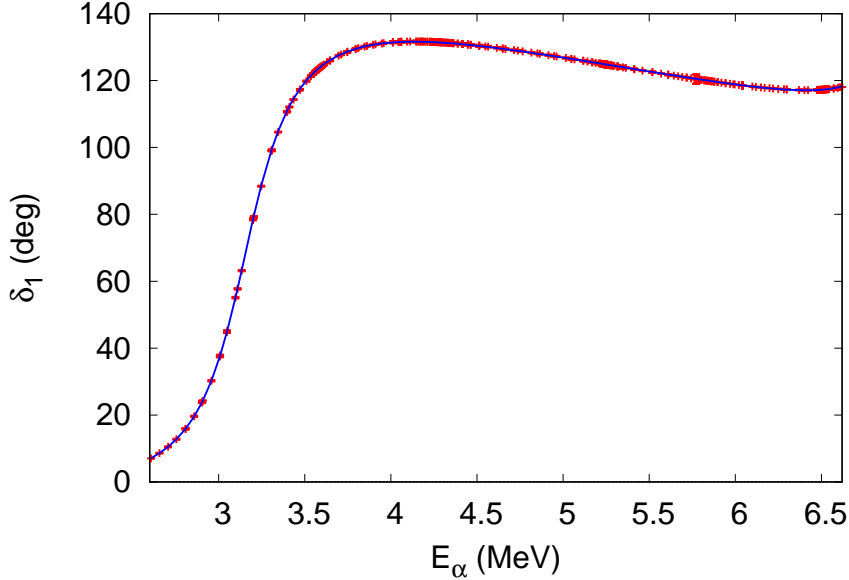}
\caption{
Phase shift $\delta_1$ of elastic $\alpha$-$^{12}$C scattering
for $p$-wave channel 
as a function of 
$E_\alpha$ calculated by using the fitted values of the parameters. 
Experimental data are included in the figure as well. 
}
\label{fig;del1}       
\end{center}
\end{figure}
In Fig.~\ref{fig;del1}, we plot a curve of the phase shift for $l=1$, 
$\delta_1$, by using the fitted values of the parameters obtained in 
Table~\ref{table;parameters_for_l=0123456}. The phase shift data are also
displayed in the figure.
We find that the fitted curve agrees well with the phase shift data.  
As discussed in Ref.~\cite{a-prc19}, the phase shift data up to 
$E_\alpha=6$~MeV can be described by the three effective range parameters
only, and the tail of the phase shift data at the high-energy side 
is now described
by the background contribution of the resonant $1_3^-$ state of $^{16}$O
from high energy.

\vskip 2mm \noindent
{\bf 4.3  Phase shift for $l=2$ channel}

We include the $2_1^+$, $2_2^+$, $2_3^+$, $2_4^+$ states of $^{16}$O
to construct an $S$ matrix for $l=2$ where 
$2_1^+$ is the sub-threshold bound state, 
$2_2^+$ and $2_3^+$ are the resonant states appearing in the phase shift data 
at $E_\alpha= 3.58$~MeV and $E_\alpha= 5.81$~MeV, respectively, 
and $2_4^+$ is a resonant state as a background contribution from high energy
appearing at $E_\alpha = 7.81$~MeV.  
Thus, we have an expression of the $S$ matrix for $l=2$ as
\bea
e^{2i\delta_2} &=& 
\frac{K_2(p) +2\kappa Re H_2(p) + ip C_\eta^2W_2(p)}
     {K_2(p) +2\kappa Re H_2(p) - ip C_\eta^2W_2(p)}
\nnb \\ && \times 
\prod_{i=2}^4
\frac{E-E_{R(2i)} + R_{(2i)}(E) - i\frac12\Gamma_{(2i)}(E)}
     {E-E_{R(2i)} + R_{(2i)}(E) + i\frac12\Gamma_{(2i)}(E)}
\,,
\label{eq;exp2idel2}
\eea
with
\bea
K_2(p) &=& 
\frac12 r_2(\gamma_2^2 + p^2)
+ \frac14P_2(\gamma_2^4 - p^4)
+ Q_2(\gamma_2^6 + p^6)
+ 2\kappa H_2(i\gamma_0)\,,
\\
\Gamma_{(2i)}(E) &=& \Gamma_{R(2i)}
\frac{pC_\eta^2W_2(p)}
     {p_rC_{\eta_r}^2W_2(p_r)}
\,,
\\
R_{(2i)}(E) &=& 
a_{(2i)} (E - E_{R(2i)})^2
+ b_{(2i)} (E - E_{R(2i)})^3\,,
\ \ \ i=2,3,4
\eea
where $\gamma_2=\sqrt{2\mu B_2}$ and $p_r =\sqrt{2\mu E_{(2i)}}$.
We have 9 parameters to fit the data as
\bea
\theta_2 &=& \{
r_2, P_2, Q_2,
E_{R(22)}, \Gamma_{R(22)}, E_{R(23)}, \Gamma_{R(23)},
a_{(23)}, b_{(23)}
\}\,,
\eea
where four parameters, $a_{(22)}$, $b_{(22)}$, $a_{(24)}$, $b_{(24)}$,
are set to be zero, $a_{(22)} = b_{(22)} = a_{(24)} = b_{(24)} = 0$, 
because they are not sensitive to the parameter fit while
we use the experimental values for $E_{R(24)}$ and $\Gamma_{R(24)}$,
$E_{R(24)}^{(exp)} = 5.858(10)$~MeV and 
$\Gamma_{R(24)}^{(exp)} =150(10)$~keV~\cite{twc-npa93},
for the background contribution from high energy. 

In Table~\ref{table;parameters_for_l=0123456},
we show fitted values and their errors of the 9 parameters in the amplitudes
of the $2_1^+$, $2_2^+$, $2_3^+$, $2_4^+$ states of $^{16}$O 
in the $S$ matrix of 
elastic $\alpha$-$^{12}$C scattering for $l=2$ in Eq.~(\ref{eq;exp2idel2})
where we 
find a small value of $\chi^2/N$, $\chi^2/N = 0.66$, for the 
parameter fit, as shown in Table~\ref{table;chi2_over_N}. 
(Those values in the table are taken from Table 2 in Ref.~\cite{a-prc22}.) 
The fitted values of $E_{R(22)}$, $\Gamma_{R(22)}$,
$E_{R(23)}$, and $\Gamma_{R(23)}$ agree with 
their experimental values, 
$E_{R(22)}^{(exp)} = 2.68255(50)$~MeV,  
$\Gamma_{R(22)}^{(exp)} = 0.625(100)$~keV,
$E_{R(23)}^{(exp)} = 4.358(4)$~MeV,  
$\Gamma_{R(23)}^{(exp)} = 71(3)$~keV~\cite{twc-npa93} while the fitted values
of the parameters, such as $Q_2$, $a_{(23)}$, and $b_{(23)}$, 
at higher order have large errors.  
In Table~\ref{table;ANC}, we show the value of the ANC of the sub-threshold 
$2_1^+$ state of $^{16}$O; we have $|C_b|=3.1(6)\times 10^4~\textrm{fm}^{-1/2}$,
which is larger than our previous estimate, 
$|C_b|=(2.1\textrm{--}2.4)\times 10^4~\textrm{fm}^{-1/2}$~\cite{a-prc18}. 
In addition, we also argued that the value of the ANC of the $2_1^+$ state of 
$^{16}$O is sensitive to conditions imposed on the effective range parameters,
$r_2$, $P_2$, $Q_2$, at very low energy region, 
$0\le E_\alpha \le 2.6~\textrm{MeV}$, where no phase shift data are reported;
we can even reproduce the large ANC value, 
$|C_b|\simeq 10\times 10^4~\textrm{fm}^{-1/2}$ reported from the $\alpha$ 
transfer experiments~\cite{bgkv-prl99,betal-npa07,ab-plb09,aetal-prl15}. 
For more details, refer to Ref.~\cite{a-prc22}. 
\begin{figure}[t]
\begin{center}
  \includegraphics[width=12cm]{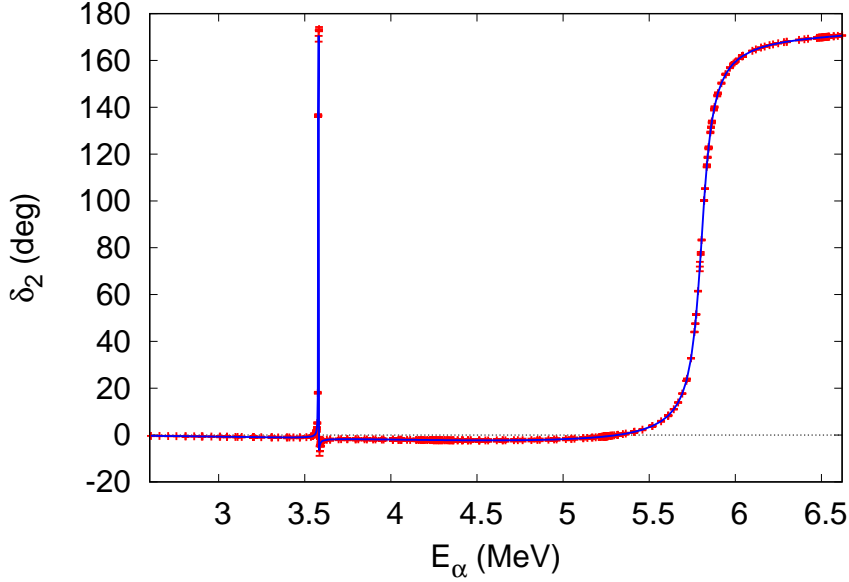}
\caption{
Phase shift $\delta_2$ of elastic $\alpha$-$^{12}$C scattering
for $d$-wave channel 
as a function of 
$E_\alpha$ calculated by using the fitted values of the parameters. 
Experimental data are included in the figure as well. 
}
\label{fig;del2}       
\end{center}
\end{figure}
In Fig.~\ref{fig;del2}, we plot a curve of the phase shift for $l=2$, 
$\delta_2$, by using the fitted values of the parameters obtained in 
Table~\ref{table;parameters_for_l=0123456}. The phase shift data are also
displayed in the figure.
We find that the fitted curve agrees well with the phase shift data.

\vskip 2mm \noindent
{\bf 4.4  Phase shift for $l=3$ channel}

We consider $3_1^-$, $3_2^-$, $3_3^-$ states of $^{16}$O 
to construct an $S$ matrix of elastic $\alpha$-$^{12}$C scattering
for $l=3$ where 
$3_1^-$ is the sub-threshold bound state, 
$3_2^-$ is the resonant state appearing in the phase shift data 
at $E_\alpha= 5.92$~MeV 
and $3_3^-$ is a resonant state as a background contribution from high energy
appearing at $E_\alpha = 7.96$~MeV.  
Because the resonant $3_2^-$ state can be described by the 
effective range parameters as well, as we have seen in the case of $l=1$, 
we include the $3_2^-$ state in the amplitude of the 
sub-threshold $3_1^-$ state.
Thus, we have an expression of the $S$ matrix for $l=3$ as
\bea
e^{2i\delta_3} &=& 
\frac{K_3(p) -2\kappa Re H_3(p) + ipC_\eta^2 W_3(p)}
     {K_3(p) -2\kappa Re H_3(p) - ipC_\eta^2 W_3(p)}
\frac{E - E_{R(33)} + R_{(33)}(E) - i\frac12\Gamma_{(33)}(E)}
     {E - E_{R(33)} + R_{(33)}(E) + i\frac12\Gamma_{(33)}(E)}
\,,
\label{eq;exp2idel3}
\eea
with
\bea
K_3(p) &=& 
\frac12r_3(\gamma_3^2 + p^2)
+ \frac14P_3(\gamma_3^4 - p^4)
+ Q_3 (\gamma_3^6 + p^6)
+ R_3 (\gamma_3^8 - p^8)
\nnb \\ &&
+ 2\kappa H_3(i\gamma_3)\,,
\\
\Gamma_{(33)}(E) &=& \Gamma_{R(33)} 
\frac{pC_\eta^2 W_3(p)}
     {p_rC_{\eta_r}^2 W_3(p_r)}
\,,
\\
R_{(33)}(E) &=& 
a_{(33)}(E - E_{R(33)})^2 
+ b_{(33)}(E - E_{R(33)})^3\,, 
\eea
where $\gamma_3 = \sqrt{2\mu B_3}$ and $p_r = \sqrt{2\mu E_{R(33)}}$.
We have 6 parameters to fit the data as
\bea
\theta_3 = \{
r_3, P_3, Q_3, R_3, a_{(33)}, b_{(33)}
\}\,,
\eea
where we impose a condition to the parameter $R_3$ for the parameter fit,
$\tilde{R}_3 < R_3$; $\tilde{R}_3$ is the contribution from the Coulomb
self-energy term, $-2\kappa H_3(p)$, to the $R_3$ term, and we have 
$\tilde{R}_3 = -17101/(90720 \kappa)$~\cite{a-prc18}.
When $\tilde{R}_3 > R_3$, a spurious bound state appears below the 
sub-threshold $3_1^-$ state of $^{16}$O. In addition, we 
use the experimental values for $E_{R(33)}$ and $\Gamma_{R(33)}$,
$E_{R(33)}^{(exp)} = 5.967(10)$~MeV and 
$\Gamma_{R(33)}^{(exp)} =110(30)$~keV~\cite{twc-npa93},
for the background contribution from high energy. 
In Table~\ref{table;parameters_for_l=0123456},
we show fitted values and their errors of the 6 parameters in the amplitudes
of the $3_1^+$, $3_2^-$, $3_3^-$ states of $^{16}$O 
in the $S$ matrix of 
elastic $\alpha$-$^{12}$C scattering for $l=3$ in Eq.~(\ref{eq;exp2idel3})
where we find a value of $\chi^2/N$, $\chi^2/N = 0.87$, for the 
parameter fit, as shown in Table~\ref{table;chi2_over_N}. 
The fitted values of the parameters $a_{(33)}$ and $b_{(33)}$ have 
large error bars and are not still fitted well because they are insensitive
in the parameter fit. While, if we exclude them from the parameter fit,
we have a large value of $\chi^2/N$, $\chi^2/N=1.84$. 
In Table~\ref{table;ANC}, we show a value of the ANC of the $3_1^-$ state
of $^{16}$O; we have $|C_b| = 113(8)~\textrm{fm}^{-1/2}$ which agrees well
with our previous estimate, 
$|C_b| = (1.2 \textrm{--} 1.5)\times 10^2~\textrm{fm}^{-1/2}$~\cite{a-prc18}.
\begin{figure}[t]
\begin{center}
  \includegraphics[width=12cm]{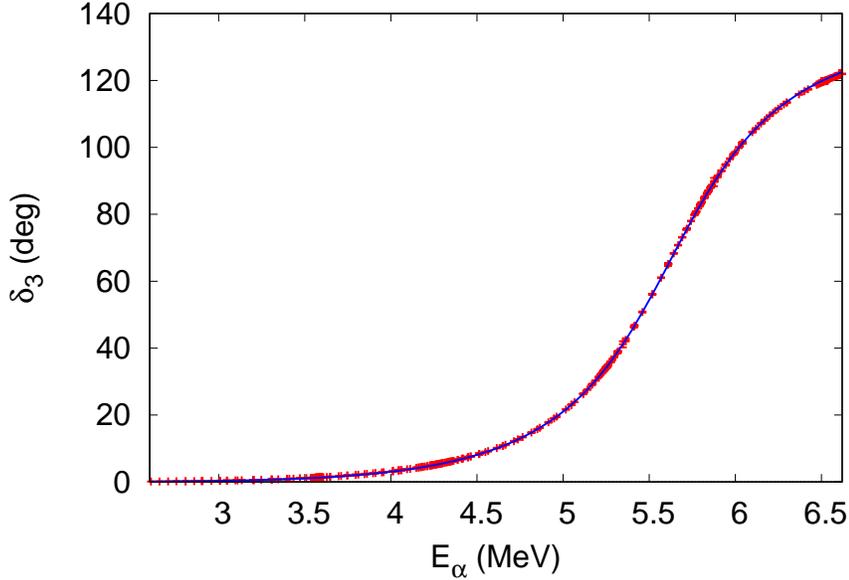}
\caption{
Phase shift $\delta_3$ of elastic $\alpha$-$^{12}$C scattering
for $f$-wave channel 
as a function of 
$E_\alpha$ calculated by using the fitted values of the parameters. 
Experimental data are included in the figure as well. 
}
\label{fig;del3}       
\end{center}
\end{figure}
In Fig.~\ref{fig;del3}, we plot a curve of the phase shift for $l=3$, 
$\delta_3$, by using the fitted values of the parameters obtained in 
Table~\ref{table;parameters_for_l=0123456}. The phase shift data are also
displayed in the figure.
We find that the fitted curve agrees well with the phase shift data.

\vskip 2mm \noindent
{\bf 4.5  Phase shift for $l=4$ channel}

We consider $4_1^+$, $4_2^+$, $4_3^+$ states of $^{16}$O 
to construct the $S$ matrix of elastic $\alpha$-$^{12}$C scattering 
for $l=4$ where
$4_1^+$ and $4_2^+$ are the resonant states appearing in the phase shift data 
at $E_\alpha= 4.26$~MeV and $E_\alpha = 5.25$~MeV, respectively, 
and $4_3^+$ is a resonant state as a background state from high energy
appearing at $E_\alpha = 8.94$~MeV.  
Thus, we have an expression of the $S$ matrix for $l=4$ as 
\bea
e^{2i\delta_4} &=& 
\prod_{i=1}^3
\frac{E-E_{R(4i)} + R_{(4i)}(E) - i\frac12\Gamma_{(4i)}(E)}
     {E-E_{R(4i)} + R_{(4i)}(E) + i\frac12\Gamma_{(4i)}(E)}
\,,
\label{eq;exp2idel4}
\eea
with
\bea
\Gamma_{(4i)}(E) &=& \Gamma_{R(4i)}
\frac{pC_\eta^2W_4(p)}
     {p_rC_{\eta_r}^2W_4(p_r)}
\,,
\\
R_{(4i)}(E) &=& 
a_{(4i)} (E - E_{R(4i)})^2
+ b_{(4i)} (E - E_{R(4i)})^3\,,
\ \ \ i=1,2,3
\eea
where 
$p_r =\sqrt{2\mu E_{(4i)}}$.
We have 8 parameters to fit to the data as
\bea
\theta_4 &=& \{
E_{R(41)}, \Gamma_{R(41)}, 
a_{(41)}, b_{(41)},
E_{R(42)}, \Gamma_{R(42)},
a_{(43)}, b_{(43)}
\}\,,
\eea
where two parameters $a_{(42)}$ and $b_{(42)}$ are set to be zero, 
$a_{(42)}=b_{(42)} = 0$, because of their insensitivity for the parameter fit,
and we use the experimental values for $E_{R(43)}$ and $\Gamma_{R(43)}$,
$E_{R(33)}^{(exp)} = 6.707(20)$~MeV and 
$\Gamma_{R(33)}^{(exp)} =89(2)$~keV~\cite{twc-npa93},
for the background contribution from high energy.

In Table~\ref{table;parameters_for_l=0123456},
we show fitted values and their errors of the 8 parameters in the amplitudes
of the $4_1^+$, $4_2^+$, $4_3^+$ states of $^{16}$O 
in the $S$ matrix of 
elastic $\alpha$-$^{12}$C scattering for $l=4$ in Eq.~(\ref{eq;exp2idel4})
where we 
find a small value of $\chi^2/N$, $\chi^2/N = 0.47$, for the 
parameter fit, as shown in Table~\ref{table;chi2_over_N}.
In addition,
the fitted values of $E_{R(41)}$, $\Gamma_{R(41)}$,
$E_{R(42)}$, and $\Gamma_{R(42)}$ agree with 
their experimental values, 
$E_{R(41)}^{(exp)} = 3.194(3)$~MeV,  
$\Gamma_{R(41)}^{(exp)} = 26(3)$~keV,
$E_{R(42)}^{(exp)} = 3.9347(17)$~MeV,  
and $\Gamma_{R(42)}^{(exp)} = 0.28(5)$~keV~\cite{twc-npa93}.
\begin{figure}[t]
\begin{center}
  \includegraphics[width=12cm]{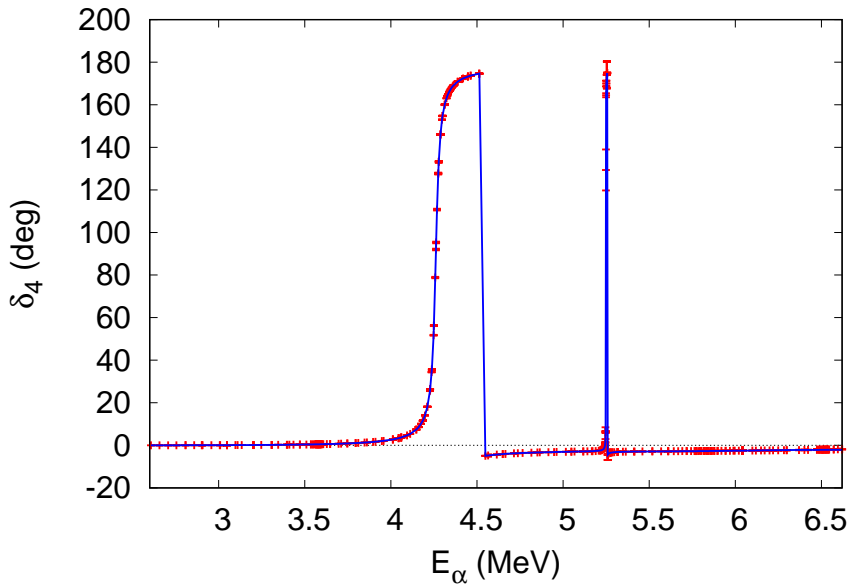}
\caption{
Phase shift $\delta_4$ of elastic $\alpha$-$^{12}$C scattering
for $g$-wave channel 
as a function of 
$E_\alpha$ calculated by using the fitted values of the parameters. 
Experimental data are included in the figure as well. 
}
\label{fig;del4}       
\end{center}
\end{figure}
In Fig.~\ref{fig;del4}, we plot a curve of the phase shift for $l=4$, 
$\delta_4$, by using the fitted values of the parameters obtained in 
Table~\ref{table;parameters_for_l=0123456}. The phase shift data are also
displayed in the figure.
We find that the fitted curve agrees well with the phase shift data.

\vskip 2mm \noindent
{\bf 4.6  Phase shift for $l=5$ channel}

We consider $5_1^-$ state of $^{16}$O 
to construct an $S$ matrix of the elastic $\alpha$-$^{12}$C scattering
for $l=5$ where 
$5_1^-$ is a resonant state as a background contribution from high energy
appearing at $E_\alpha = 10.00$~MeV.  
Thus, we have an expression of the $S$ matrix for $l=5$ as 
\bea
e^{2i\delta_5} &=& 
\frac{E-E_{R(51)} + R_{(51)}(E) - i\frac12\Gamma_{(51)}(E)}
     {E-E_{R(51)} + R_{(51)}(E) + i\frac12\Gamma_{(51)}(E)}
\,,
\label{eq;exp2idel5}
\eea
with
\bea
\Gamma_{(51)}(E) &=& \Gamma_{R(51)}
\frac{pC_\eta^2W_5(p)}
     {p_rC_{\eta_r}^2W_5(p_r)}
\,,
\\
R_{(51)}(E) &=& 
a_{(51)} (E - E_{R(51)})^2
+ b_{(51)} (E - E_{R(51)})^3\,,
\eea
where $p_r =\sqrt{2\mu E_{(51)}}$.
We have 2 parameters to fit the data as
\bea
\theta_4 &=& \{
a_{(51)}, b_{(51)}
\}\,,
\eea
where we use the experimental values for $E_{R(51)}$ and $\Gamma_{R(51)}$,
$E_{R(51)}^{(exp)} = 7.498(2)$~MeV and 
$\Gamma_{R(51)}^{(exp)} =670(15)$~keV~\cite{twc-npa93},
for the background contribution from high energy.

In Table~\ref{table;parameters_for_l=0123456},
we show fitted values and their errors of the 2 parameters in the amplitudes
of the $5_1^-$ state of $^{16}$O 
in the $S$ matrix of 
elastic $\alpha$-$^{12}$C scattering for $l=5$ in Eq.~(\ref{eq;exp2idel5})
where we find a very small value of $\chi^2/N$, $\chi^2/N = 0.094$, for the 
parameter fit, as shown in Table~\ref{table;chi2_over_N}.
\begin{figure}[t]
\begin{center}
  \includegraphics[width=12cm]{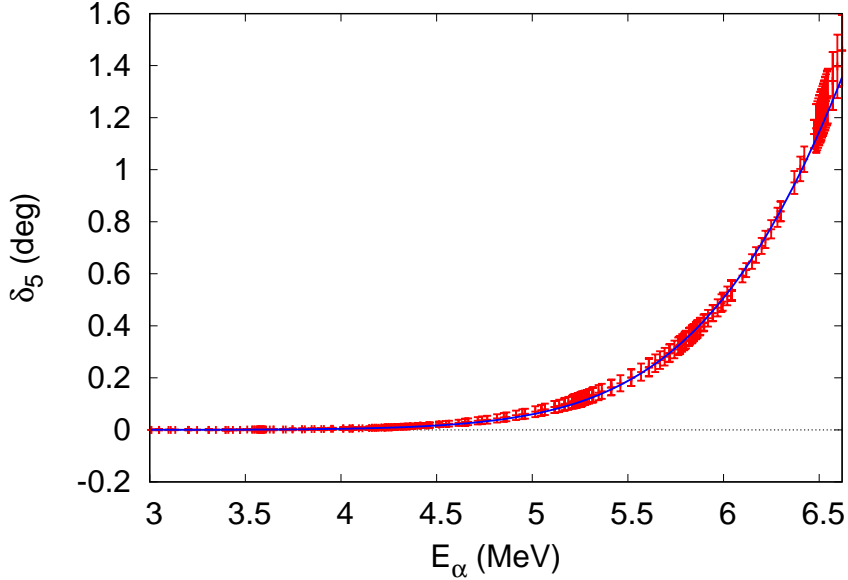}
\caption{
Phase shift $\delta_5$ of elastic $\alpha$-$^{12}$C scattering
for $h$-wave channel 
as a function of 
$E_\alpha$ calculated by using the fitted values of the parameters. 
Experimental data are included in the figure as well. 
}
\label{fig;del5}       
\end{center}
\end{figure}
In Fig.~\ref{fig;del5}, we plot a curve of the phase shift for $l=5$, 
$\delta_5$, by using the fitted values of the parameters obtained in 
Table~\ref{table;parameters_for_l=0123456}. The phase shift data are also
displayed in the figure.
We find that the fitted curve agrees well with the phase shift data.

\vskip 2mm \noindent
{\bf 4.7  Phase shift for $l=6$ channel}

We include a background contribution from low energy
and the $6_1^+$ state of $^{16}$O
to construct an $S$ matrix of the elastic $\alpha$-$^{12}$C scattering
for $l=6$ where
$6_1^+$ is a resonant state as a background contribution from high energy
appearing at $E_\alpha = 10.20$~MeV.  
Thus, we have an expression of the $S$ matrix for $l=6$ as 
\bea
e^{2i\delta_6} &=& 
\frac{K_6(p) -2\kappa Re H_6(p) + ipC_\eta^2 W_6(p)}
     {K_6(p) -2\kappa Re H_6(p) - ipC_\eta^2 W_6(p)}
\frac{E - E_{R(61)} + R_{(61)}(E) - i\frac12\Gamma_{(61)}(E)}
     {E - E_{R(61)} + R_{(61)}(E) + i\frac12\Gamma_{(61)}(E)}
\,,
\label{eq;exp2idel6}
\eea
with
\bea
K_6(p) &=& 
\frac12r_6 p^2
- \frac14P_6 p^4
\,,
\\
\Gamma_{(61)}(E) &=& \Gamma_{R(61)} 
\frac{pC_\eta^2 W_6(p)}
     {p_rC_{\eta_r}^2 W_6(p_r)}
\,,
\\
R_{(61)}(E) &=& 
a_{(61)}(E - E_{R(61)})^2 
+ b_{(61)}(E - E_{R(61)})^3\,, 
\eea
where $p_r = \sqrt{2\mu E_{R(61)}}$.
We have four parameters to fit the data as
\bea
\theta_3 = \{
r_6, P_6, a_{(61)}, b_{(61)} 
\}\,,
\eea
where the leading effective range parameter, $-1/a_6$, is put to be zero,
$1/a_6=0$, and two effective range parameters, $r_6$ and $P_6$, are included
while we use the experimental values for $E_{R(61)}$ and $\Gamma_{R(61)}$,
$E_{R(61)}^{(exp)} = 7.6534(16)$~MeV and 
$\Gamma_{R(61)}^{(exp)} =70(8)$~keV~\cite{twc-npa93},
for the background contribution from high energy.

In Table~\ref{table;parameters_for_l=0123456},
we show fitted values and their errors of the four parameters in the amplitudes
of the background contribution from low energy and 
the $6_1^+$ state of $^{16}$O in the $S$ matrix of 
elastic $\alpha$-$^{12}$C scattering for $l=6$ in Eq.~(\ref{eq;exp2idel6})
where we 
find a very small value of $\chi^2/N$, $\chi^2/N = 0.026$, for the 
parameter fit, as shown in Table~\ref{table;chi2_over_N}.
\begin{figure}[t]
\begin{center}
  \includegraphics[width=12cm]{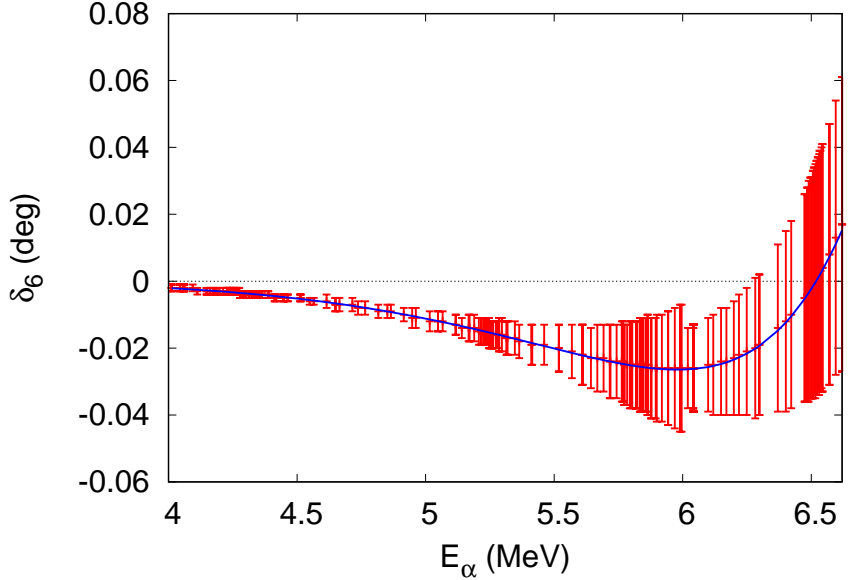}
\caption{
Phase shift $\delta_6$ of elastic $\alpha$-$^{12}$C scattering
for $i$-wave channel 
as a function of 
$E_\alpha$ calculated by using the fitted values of the parameters. 
Experimental data are included in the figure as well. 
}
\label{fig;del6}       
\end{center}
\end{figure}
In Fig.~\ref{fig;del6}, we plot a curve of the phase shift for $l=6$, 
$\delta_6$, by using the fitted values of the parameters obtained in 
Table~\ref{table;parameters_for_l=0123456}. The phase shift data are also
displayed in the figure.
We find that the fitted curve agrees well with the phase shift data.

\vskip 2mm 
\noindent 
{\bf 5. Results and discussion}

In the present work, we studied the expression of the $S$ matrices of 
elastic $\alpha$-$^{12}$C scattering at low energies for $l=0,1,2,3,4,5,6$
in EFT. The $S$ matrices are constructed as the summation of the 
phase shifts; the parts of the phase shifts are obtained from the 
elastic scattering amplitudes of the sub-threshold and resonant states
of $^{16}$O. Those amplitudes are calculated from the effective Lagrangian
and obtained in terms of the effective range parameters up to $p^6$ 
order for $l=0,1,2,4,5,6$ and up to $p^8$ order for $l=3$ due to the 
modification of the counting rules for the effective range parameters
discussed in Refs.~\cite{a-prc19,a-prc22}. 
We include the sub-threshold states for $l=0,1,2,3$ and a background 
contribution from low energy for $l=6$, 
one resonant state for $l=0,1,3$ and two resonant states for $l=2,4$,
which appear in the energy range of the phase shift data,
$2.6~\textrm{MeV} < E_\alpha < 6.62~\textrm{MeV}$, and a resonant state
which appears at $6.62~\textrm{MeV} < E_\alpha$,
as background contributions from high energy 
for all the partial wave states. 
Those states included in
the present study are summarized in Table~\ref{table;states}.
Then, we fit the parameters to the phase shift data 
while the resonant energies and widths of the background contributions
from high energy are fixed by
using the experimental data of the resonant states 
at $6.62~\textrm{MeV} < E_\alpha$, and some parameters which are insensitive to 
the parameter fit are suppressed. We find that the parameters are fitted
very well where the $\chi^2/N$ values are less than one for all the cases,
as summarized in Table~\ref{table;chi2_over_N},
and thus, the phase shifts are well described within the theory.  

One may regard this work as merely a simple parameter fit to the 
phase shift data while our aim is to extrapolate the radiative capture rate
down to the Gamow peak energy, $E_G = 0.3~\textrm{MeV}$;
one may see that our result is comparable to that
worked out by the $R$-matrix analysis. One may also see that the 
parameterization based on the effective range expansion in the $S$ matrices
in Eq.~(\ref{eq;exp2idel_l}) is simple and transparent. In addition,
the electromagnetic and weak interactions are straightforwardly 
introduced in the theory. 
As mentioned in the introduction, we employed  
the EFT to the study of the $E1$ transition of 
the $^{12}$C($\alpha$,$\gamma$)$^{16}$O reaction~\cite{a-prc19} 
and the $\beta$-delayed $\alpha$-emission 
from $^{16}$N~\cite{a-epja21}.\footnote{
The dressed propagators of $^{16}$O 
shown in Fig.~\ref{fig;propagator},
$D_l^{-1}(p)$ for $l=1$ and $l=3$,
whose parameters are fitted to the phase shift data, are
used as the building blocks of reaction amplitudes of the $E1$ transition
of $^{12}$C($\alpha$,$\gamma$)$^{16}$O and the $\beta$-delayed $\alpha$-emission
of $^{16}$N. 
}
Recently, we reported the first application
of EFT to the study of the radiative proton capture 
on $^{15}$N~\cite{sao-SMNP22,sao-22}, which is
an important reaction in the CNO cycle. Thus, constructing an EFT would 
be another theoretical method, 
as an alternative to the $R$-matrix analysis, 
for the studies of nuclear reactions for stellar evolution.

\vskip 2mm \noindent
{\bf Acknowledgements}

We would like to thank S.~W. Hong, T.-S. Park, and C.~H. Hyun 
for discussions. 
We also thank the reviewers for the discussions and for providing 
the data files of wavefunctions and phase shifts calculated from
a Woods-Saxon potential. 
This work was supported by 
the National Research Foundation of Korea (NRF) grant funded by the 
Korean government (MSIT) (No. 2019R1F1A1040362 and 2022R1F1A1070060). 

\vskip 2mm \noindent
{\bf Appendix}

In the present appendix, we perform a test calculation by employing
a Woods-Saxon potential. 
The data, wavefunctions and phase shifts, are generated by using 
the model potential, the values of ANCs are calculated 
from the wavefunctions, and
the scattering phase shifts are generated by using the potential.
We, then, fit the effective range parameters to the phase shift data
generated from the potential, and calculate the ANCs by using 
the fitted values of parameters in Eq.~(\ref{eq;Cb}).  
We note that the relation between the two methods, the potential model
and the effective range expansion, to deduce the ANCs is not obvious. 
The comparison of the ANCs from the potential model and the fit of 
effective range parameters may indicate some degree of the model 
dependence in the two methods. 

\begin{figure}[t]
\begin{center}
  \includegraphics[width=12cm]{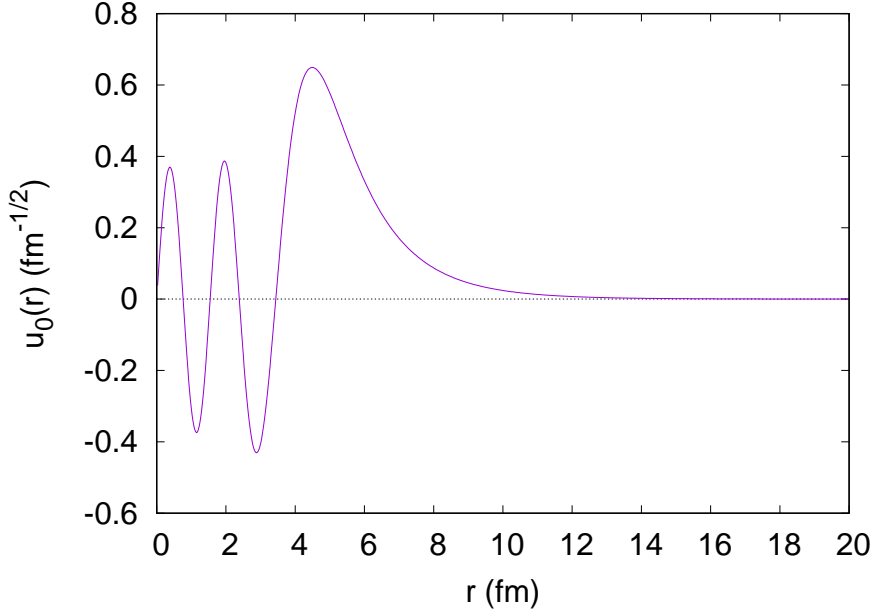}
\caption{Wavefunction for $0^+$ state of $^{16}$O calculated 
by using a Woods-Saxon potential, which  
is normalized as $\int_0^\infty u_0(r)^2 dr = 1$.
}
\label{fig;wf0}       
\end{center}
\end{figure}
\begin{figure}[t]
\begin{center}
  \includegraphics[width=12cm]{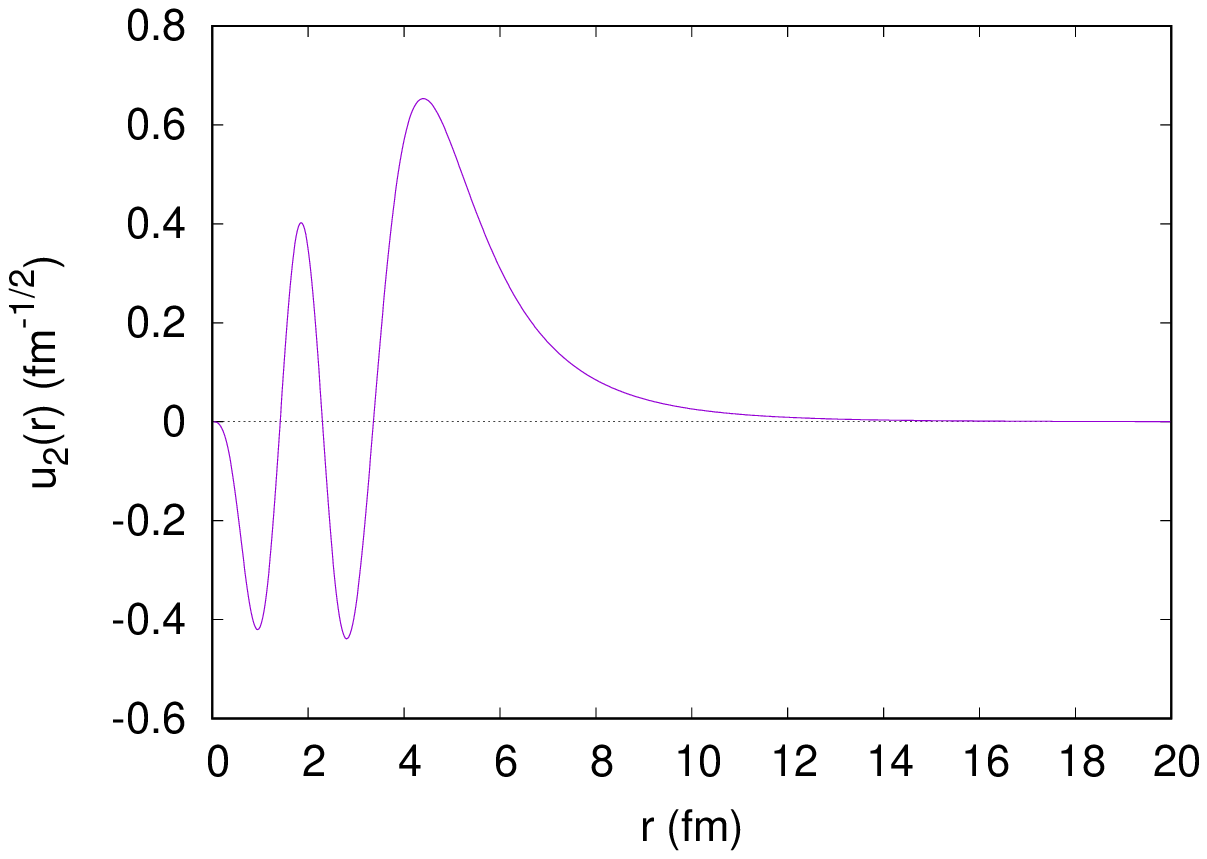}
\caption{Wavefunction for $2^+$ state of $^{16}$O calculated 
by using a Woods-Saxon potential, which  
is normalized as $\int_0^\infty u_2(r)^2 dr = 1$.
}
\label{fig;wf2}       
\end{center}
\end{figure}
In Figs.~\ref{fig;wf0} and \ref{fig;wf2}, we display the wavefunctions
of the $0^+$ and $2^+$ states of $^{16}$O, respectively, 
from a Woods-Saxon potential.
The geometry of Woods-Saxon potential is standard; 
$R_0=1.3$~fm; $R=R_0 A^{1/3} = 2.976$~fm and $a=0.7$~fm. 
The wavefunctions of bound states are obtained by adjusting the depth of
potential. For the $0^+$ state, $E_x = 6049.4$~keV and $B_0=1112.5$~keV, 
and one has $V_0 = 131.1$~MeV, and for the $2^+$ state, $E_x = 6917.1$~keV
and $B_2 = 244.8$~keV, and one has $V_0=130.8$~MeV. 
The wavefunctions are normalized as $\int_0^\infty u_{l}(r)^2dr = 1$ 
with $l=0,2$. 
(The data files of wavefunctions and phase shifts are provided by a reviewer.) 

The values of ANCs may be obtained by using the relations
\bea
u_0(r) \sim |C_b|_0 W_{-\kappa/\gamma_0, \frac12}(2\gamma_0r)\,,
\ \ \ \ 
u_2(r) \sim |C_b|_2 W_{-\kappa/\gamma_2, \frac52}(2\gamma_2r)\,,
\label{eq;ANCs_WS}
\eea
at the outside of the potential range, $R < r$, 
where $W_{\nu,\mu}(z)$ is the Whittaker function, and 
$\gamma_0$ and $\gamma_2$ are the binding momenta,   
$\gamma_0 =\sqrt{2\mu B_0}$ and 
$\gamma_2 =\sqrt{2\mu B_2}$. Thus, we have the ANCs from the wavefunctions
as
\bea
|C_b|_0 = 2.6 \times 10^3 \ \ \textrm{fm}^{-1/2}\,,
\ \ \ \
|C_b|_2 = 1.9 \times 10^5 \ \ \textrm{fm}^{-1/2}\,.
\eea 
It is good to see that the ANCs are straightforwardly obtained from
the wavefunctions generated from the potential model after fitting
the depth of potential to the binding energies. 
While one may wonder about the sensitivity of ANCs to all the three parameters
of potential and the number of nodes of wavefunctions inside the potential
range. In addition, the ANCs from the potential model depend on 
the normalization condition of wavefunctions; the reliability of two-body 
description of the 16-body nucleon system inside the potential range 
is questionable. 

\begin{figure}[t]
\begin{center}
  \includegraphics[width=12cm]{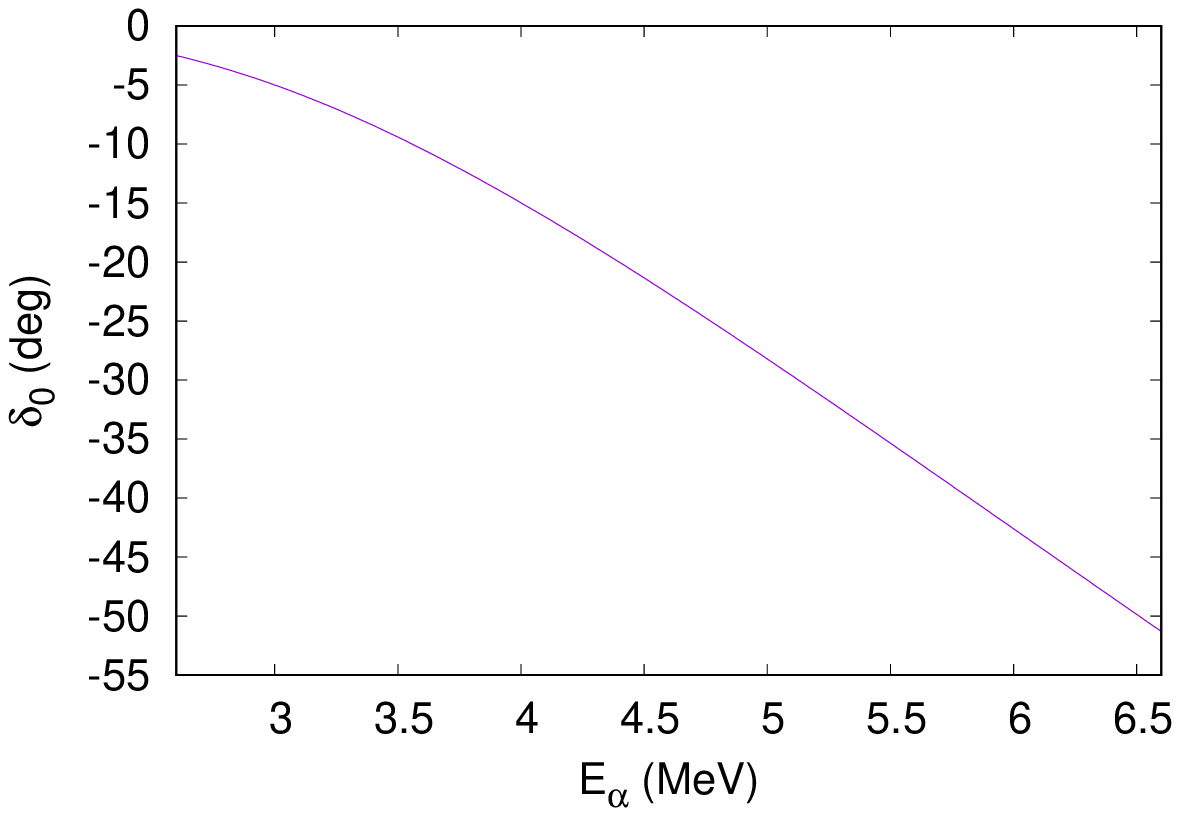}
\caption{
Phase shift $\delta_0$ of elastic $\alpha$-$^{12}$C scattering for $l=0$ 
as a function of the $\alpha$ energy $E_\alpha$ in the lab mass frame,
calculated by using a Woods-Saxon potential. 
}
\label{fig;phase0}       
\end{center}
\end{figure}
\begin{figure}[t]
\begin{center}
  \includegraphics[width=12cm]{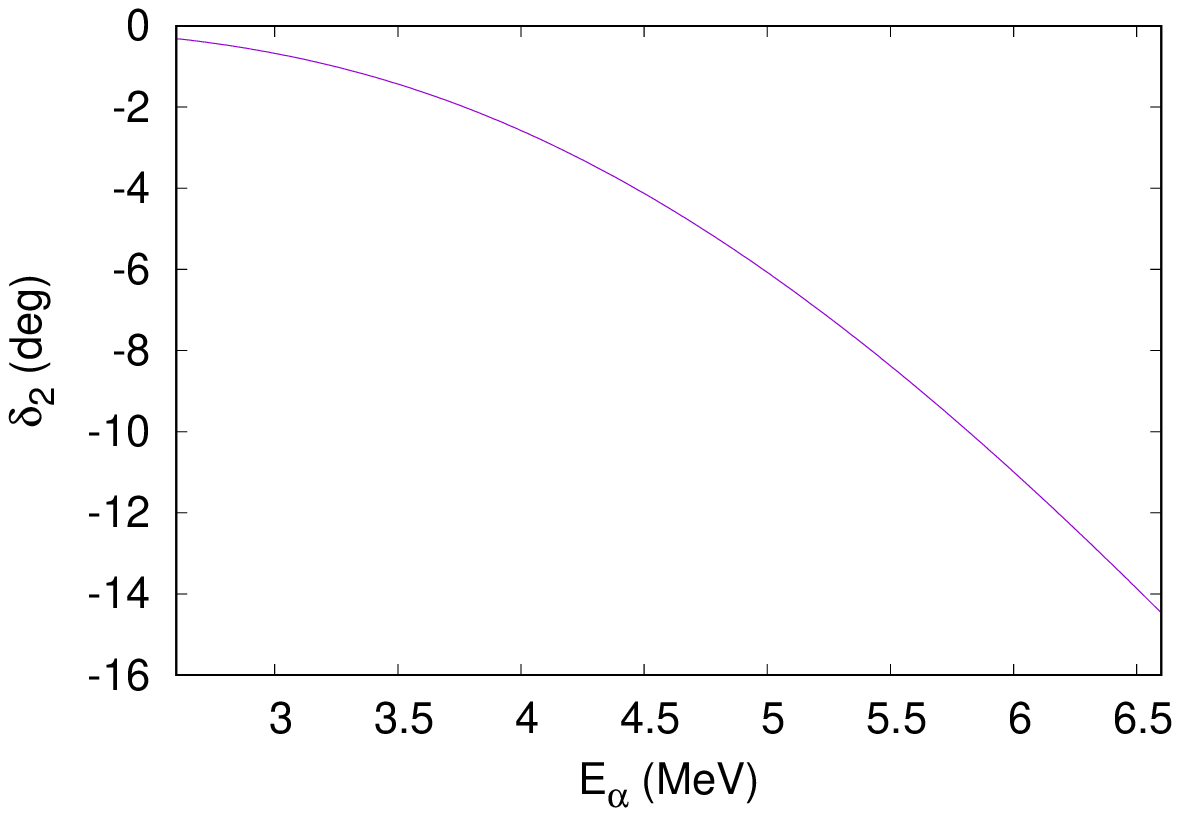}
\caption{
Phase shift $\delta_2$ of elastic $\alpha$-$^{12}$C scattering for $l=2$ 
as a function of the $\alpha$ energy $E_\alpha$ in the lab frame,
calculated by using a Woods-Saxon potential. 
}
\label{fig;phase2}       
\end{center}
\end{figure}
In Figs.~\ref{fig;phase0} and \ref{fig;phase2}, we display the phase shifts
of elastic $\alpha$-$^{12}$C scattering for $l=0$ and 2, respectively,
calculated from the Woods-Saxon potential with the fixed values of $V_0$ 
mentioned above. Now we carry out a test for our fitting method to deduce 
the ANCs from the phase shift data: the effective range parameters are fitted
to the phase shift data in the figures, and we calculate the ANCs by using 
Eq.~(\ref{eq;Cb}). When fitting the phase shift data, we include a constant
error to the phase shift data, $\Delta \delta_{0,2} = 0.20$ degree, which
is a typical size of the error in the phase shift data for $l=0$ 
from the experiment. 

\begin{table}[h]
\begin{center}
\begin{tabular}{c | c c c c } 
$E_{\alpha,max}$ (MeV) & 4.0 & 4.5 & 5.0 & 5.5 
\cr \hline
$|C_b|_0$ ($\textrm{fm}^{-1/2}$) & $1.1\times 10^3$ & $2.2\times 10^3$ & 
 $3.7\times 10^3$ & $5.8\times 10^3$
\cr
$\chi^2/N$ & 0.05 & 0.16 & 0.43 & 1.18
\end{tabular}
\caption{
ANC of $0^+$ state, $|C_b|_0$, as a function of the maximum energy 
of the data, $E_{\alpha,max}$ (MeV), deduced from the phase shift data
in Fig.~\ref{fig;phase0}.
Values of $\chi^2/N$ are also displayed in the table. 
}
\label{table;ANC_0}
\end{center}
\end{table}
In Table~\ref{table;ANC_0}, we display the values of ANC for the $0^+$ state
as a function of the maximum energy, 
$E_{\alpha, max} = 4.0$, 4.5, 5.0, 5.5~MeV, of the phase shift data 
in Fig.~\ref{fig;phase0}. We find that the shapes of the curves of the phase
shift are different (the curve from the effective range parameters is 
stiffer than that from the potential model), and it leads to the energy
dependence of the fit.~\footnote{
The different energy dependence in the phase shifts possibly came out 
due to the use of an energy-independent potential model, where one may notice
that the calculated phase shifts do not agree with the experimental data.
The use of an energy-dependent potential could improve the present situation. 
} 
One can see that the values of ANC in the table have 
a significant range, $(1.1 \textrm{--} 5.8)\times 10^3$~fm$^{-1/2}$ with 
$\chi^2/N = 0.05 \textrm{--} 1.18$, while the ANC from the potential model,
$2.6\times 10^3$~fm$^{-1/2}$, can be found within the range of ANC from the
fit.   
For the $2^+$ state, using all the phase shift data in Fig.~\ref{fig;phase2},
we have 
\bea
|C_b|_2 = 3.4(54.4)\times 10^5~\textrm{fm}^{-1/2}\,,
\eea
with $\chi^2/N = 0.13$. 
The center value of ANC is 1.8 times larger than that of the potential
model in Eq.~(\ref{eq;ANCs_WS}) while the ANC from fit has a large error 
bar. As discussed in Ref.~\cite{a-prc22}, when the center value of $dD/dp^2$ 
term at $p=i\gamma_2$ almost vanishes in the denominator
in Eq.~(\ref{eq;Cb}), the error of ANC is enhanced.   

In this appendix, we performed a test calculation to study whether the ANCs
obtained from the wavefunctions of a potential model can be reproduced
by fitting the effective range parameters to the phase shift generated
by the model potential. We find that the ANCs obtained from the fitted
values of effective range parameters agree with those obtained from the 
wavefunction of the potential model within the uncertainties discussed above;
aside from the large uncertainties, the center values of ANCs agree within
a factor of 2. The large uncertainties may stem from the formula of ANC 
in Eq.~(\ref{eq;Cb}). When the slope of the inverse of the propagator, $D_l(p)$,
becomes small at the binding momentum $p = i\gamma_l$, the ANC becomes large.
In other words, the center value of ANC becomes sensitive to the values 
of effective range parameters while the errors of ANC are enhanced. 
Therefore, because of the difficulty discussed above, the present cases
to deduce the ANCs of $0^+$ and $2^+$ state from the phase shift data 
may not be ideal: the uncertainty of a factor of 2 may be involved
between the two methods.

\end{document}